\pdfoutput=1
\documentclass[11pt,twoside,a4paper,cmspaper,final,collab]{cms-tdr}
\def\svnVersion{250a077}\def\svnDate{2023/08/25}\def\cmsCernNoTag{CERN-EP-2022-223}\def\cmsCernDate{\today}\def\cmsMessage{Published in Physical Review D as \href{http://dx.doi.org/10.1103/PhysRevD.108.032017}{\doi{10.1103/PhysRevD.108.032017}.}}
\begin{document}\cmsNoteHeader{SMP-21-011}

\newlength\cmsTabSkip\setlength{\cmsTabSkip}{1ex}
\ifthenelse{\boolean{cms@external}}{\providecommand{\cmsLeft}{upper\xspace}}{\providecommand{\cmsLeft}{left\xspace}}
\ifthenelse{\boolean{cms@external}}{\providecommand{\cmsRight}{lower\xspace}}{\providecommand{\cmsRight}{right\xspace}}

\newcommand{\sdev}{\unit{SD}}
\newcommand{\MJ}{\ensuremath{m_\text{J}^\text{SD}}\xspace}
\newcommand{\Jg}{\ensuremath{\text{J}\Pgg}\xspace}
\newcommand{\Hg}{\ensuremath{\PH\Pgg}\xspace}
\newcommand{\mll}{\ensuremath{m_{\Pell\Pell}}}
\newcommand{\dyll}{\ensuremath{\PZ/\Pgg^*\to\Pell^+\Pell^-}}
\newcommand{\jj}{\ensuremath{\text{jj}}\xspace}
\newcommand{\alphahat}{\ensuremath{\hat{\alpha}}\xspace}
\newcommand{\loperator}[2]{\ensuremath{\mathcal{L}_{\text{#1,#2}}}\xspace}
\newcommand{\fcoeff}[2]{\ensuremath{f_{\text{#1,#2}}}\xspace}
\newcommand{\fcoefflam}[2]{\ensuremath{f_{\text{#1,#2}}/\Lambda^{4}}\xspace}
\newcommand{\ubound}{\ensuremath{U_{\text{bound}}}\xspace}
\newcommand{\escat}{\ensuremath{E_{\text{scattering}}}\xspace}
\newcommand{\dnll}{\ensuremath{\Delta\text{NLL}}\xspace}
\newcommand{\dejj}{\ensuremath{\Delta\etajj}\xspace}
\newcommand{\dphilmet}{\ensuremath{\Delta\phi_{\Pell,\ptmiss}}\xspace}
\newcommand{\muhat}{\ensuremath{\hat{\mu}}\xspace}
\newcommand{\fremu}{\ensuremath{f_{\Pe{}(\Pgm{})}}\xspace}
\newcommand{\frlep}{\ensuremath{f_{\Pell}}\xspace}
\newcommand{\muR}{\ensuremath{\mu_{\text{R}}}\xspace}
\newcommand{\muF}{\ensuremath{\mu_{\text{F}}}\xspace}
\newcommand{\thetab}{\ensuremath{\boldsymbol{\theta}}\xspace}
\newcommand{\thetabhat}{\ensuremath{\hat{\thetab}}\xspace}
\newcommand{\thetabdhat}{\ensuremath{\hat{\thetabhat}}\xspace}
\newcommand{\sigmag}{\ensuremath{\sigma_\text{g}}\xspace}
\newcommand{\sigmagew}{\ensuremath{\sigmag^\text{EW}}\xspace}
\newcommand{\sigmagqcd}{\ensuremath{\sigmag^\text{QCD}}\xspace}
\newcommand{\sigmaee}{\ensuremath{\sigma_{\eta\eta}}\xspace}
\newcommand{\sigfid}{\ensuremath{\sigma^{\text{fid}}}\xspace}
\newcommand{\sigfidew}{\ensuremath{\sigfid_{\text{EW}}}\xspace}
\newcommand{\sigfidewq}{\ensuremath{\sigfid_{\text{EW+QCD}}}\xspace}
\newcommand{\accgf}{\ensuremath{\text{a}_\text{gf}}\xspace}
\newcommand{\alphagf}{\ensuremath{\alpha_\text{gf}}\xspace}
\newcommand{\alphaewgf}{\ensuremath{\alphagf^\text{EW}}\xspace}
\newcommand{\alphaqcdgf}{\ensuremath{\alphagf^\text{QCD}}\xspace}
\newcommand{\alphatest}{\ensuremath{\alpha_\text{test}}\xspace}
\newcommand{\WWjj}{\ensuremath{\PW^{\pm}\PW^{\pm}\jj}\xspace}
\newcommand{\ttg}{\ensuremath{\ttbar\Pgg}\xspace}
\newcommand{\WW}{\ensuremath{\PW{}\PW}\xspace}
\newcommand{\WZ}{\ensuremath{\PW{}\PZ}\xspace}
\newcommand{\ZZ}{\ensuremath{\PZ{}\PZ}\xspace}
\newcommand{\Wg}{\ensuremath{\PW{}\Pgg}\xspace}
\newcommand{\VV}{\ensuremath{\PV{}\PV}\xspace}
\newcommand{\pp}{\ensuremath{\Pp{}\Pp}\xspace}
\newcommand{\mWg}{\ensuremath{m_{\Wg{}}}\xspace}
\newcommand{\Wgjj}{\ensuremath{\PW{}\Pgg{}\jj}\xspace}
\newcommand{\WWgg}{\ensuremath{\PW{}\PW{}\Pgg{}\Pgg}\xspace}
\newcommand{\Zg}{\ensuremath{\PZ{}\Pgg}\xspace}
\newcommand{\gpjets}{\Pgg{}+jets\xspace}
\newcommand{\Wpjets}{\PW{}+jets\xspace}
\newcommand{\Wgpjets}{\PW{}\Pgg{}+jets\xspace}
\newcommand{\Zpjets}{\PZ{}+jets\xspace}
\newcommand{\Zgpjets}{\PZ{}\Pgg{}+jets\xspace}
\newcommand{\lepg}{\ensuremath{\Pell{}\Pgg}\xspace}
\newcommand{\mjj}{\ensuremath{m_{\jj}}\xspace}
\newcommand{\mlepg}{\ensuremath{m_{\lepg}}\xspace}
\newcommand{\lnugjj}{\ensuremath{\Pell{}\Pgn{}\Pgg{}\jj}\xspace}
\newcommand{\mTW}{\ensuremath{\mT^\PW}\xspace}
\newcommand{\pz}{\ensuremath{p_{\text{z}}}\xspace}
\newcommand{\ptX}[1]{\ensuremath{\pt^{#1}}\xspace}
\newcommand{\reliso}{\ensuremath{I^{\ell}}\xspace}
\newcommand{\PUpt}{\ensuremath{\ptX{\text{PU}}}\xspace}
\newcommand{\ptg}{\ensuremath{\ptX{\Pgg}}\xspace}
\newcommand{\ptlep}{\ensuremath{\ptX{\Pell}}\xspace}
\newcommand{\ptjj}{\ensuremath{\ptX{\jj}}\xspace}
\newcommand{\ptj}{\ensuremath{\ptX{\text{j}}}\xspace}
\newcommand{\ptW}{\ensuremath{\ptX{\PW}}\xspace}
\newcommand{\ptZ}{\ensuremath{\ptX{\PZ}}\xspace}
\newcommand{\etaX}[1]{\ensuremath{\eta_{#1}}\xspace}
\newcommand{\etaSC}{\etaX{\text{SC}}}
\newcommand{\etag}{\etaX{\Pgg{}}}
\newcommand{\etajj}{\etaX{\jj}}
\newcommand{\etaj}{\etaX{\text{j}}}
\newcommand{\etaW}{\etaX{\PW{}}}
\newcommand{\etaZ}{\etaX{\PZ{}}}
\newcommand{\etalep}{\etaX{\Pell{}}}
\newcommand{\ptch}{\ensuremath{\ptX{\text{charged}}}\xspace}
\newcommand{\ptneu}{\ensuremath{\ptX{\text{neutral}}}\xspace}
\newcommand{\sumg}{\ensuremath{\sum\ptg}\xspace}
\newcommand{\sumch}{\ensuremath{\sum\ptch}\xspace}
\newcommand{\sumneu}{\ensuremath{\sum\ptneu}\xspace}
\newcommand{\phiWg}{\ensuremath{\phi_{\Wg{}}}\xspace}
\newcommand{\phiDijet}{\ensuremath{\phi_{\text{jj}}}\xspace}
\newcommand{\kinsysY}{\ensuremath{\abs{y_{\Wg{}}-(y_{\text{j1}}+y_{\text{j2}})/2}}\xspace}
\newcommand{\kinsysPhi}{\ensuremath{\abs{\phiWg - \phiDijet}}\xspace}
\newcommand{\eletogam}{\ensuremath{\Pe\to\Pgg}\xspace}

\cmsNoteHeader{SMP-21-011}

\title{Measurement of the electroweak production of \texorpdfstring{$\PW\gamma$}{W gamma} in association with two jets in proton-proton collisions at \texorpdfstring{$\sqrt{s}=13\TeV$}{sqrt(s) = 13 TeV}}

\date{\today}

\abstract{A measurement is presented for the electroweak production of a \PW boson, a photon ($\gamma$), and two jets (j) in proton-proton collisions. The leptonic decay of the \PW boson is selected by requiring one identified electron or muon and large missing transverse momentum. The two jets are required to have large invariant dijet mass and large separation in pseudorapidity. The measurement is performed with the data collected by the CMS detector at a center-of-mass energy of 13\TeV, corresponding to an integrated luminosity of 138\fbinv. The cross section for the electroweak $\PW\gamma$jj production is $23.5^{+4.9}_{-4.7}\unit{fb}$, whereas the total cross section for $\PW\gamma$jj production is $113\pm 13\unit{fb}$. Differential cross sections are also measured with the distributions unfolded to the particle level. All results are in agreement with the standard model expectations. Constraints are placed on anomalous quartic gauge couplings (aQGCs) in terms of dimension-8 effective field theory operators. These are the most stringent limits to date on the aQGCs parameters \fcoefflam{M}{2--5} and \fcoefflam{T}{6--7}.}

\hypersetup{
pdfauthor={CMS Collaboration},
pdftitle={Measurement of the electroweak production of W gamma in association with two jets in proton-proton collisions at sqrt(s) = 13 TeV},
pdfsubject={CMS},
pdfkeywords={CMS,  vector boson scattering}}

\maketitle

\section{Introduction}
\label{sec:introduction}
The discovery of the Higgs boson at the CERN LHC~\cite{Aad:2012tfa,Chatrchyan:2012ufa,Chatrchyan_2013} was made about ten years ago.
Now, it is of great interest to examine in depth the mechanism of electroweak (EW) symmetry breaking using rare EW processes.
Vector boson scattering (VBS) processes play an independent and complementary role in understanding the EW symmetry breaking. 
The nonabelian nature of gauge interactions in the standard model (SM) leads to
a large variety of VBS processes with unique features and opportunities to probe new physics beyond the SM (BSM).

The center-of-mass energy of the proton-proton(\pp) collisions and the integrated luminosity accumulated by the LHC experiments present an opportunity to measure many rare VBS processes.
For example, the observed (expected) significance for the EW production of \Wg + 2 jets reported by CMS is 5.3 (4.8) standard deviations (\sdev)
combining Run 1 data and Run 2 data collected in 2016~\cite{CMS:2020ypo}.

This paper presents a measurement of the EW \Wgjj production at $\sqrt{s}=13\TeV$ based on the complete Run 2 data collected during 2016--2018, superseding the previous CMS result~\cite{CMS:2020ypo}.
A complete set of tabulated results of this analysis is available in the HEPData database~\cite{hepdata}.
In addition to increased integrated luminosity, our new results include: ($\romannumeral1$) an updated fiducial region requiring jets with $\pt > 50\GeV$; ($\romannumeral2$) the removal of the missing transverse momentum requirement from the fiducial region definition; ($\romannumeral3$) the treatment of the interference term between the EW- and quantum chromodynamics (QCD) induced processes as a background component; ($\romannumeral4$) and the treatment of the out-of-fiducial signal contribution as a background component.

\begin{figure*}[ht!]
  \centering
    \includegraphics[height=0.21\textwidth,width=0.23\textwidth]{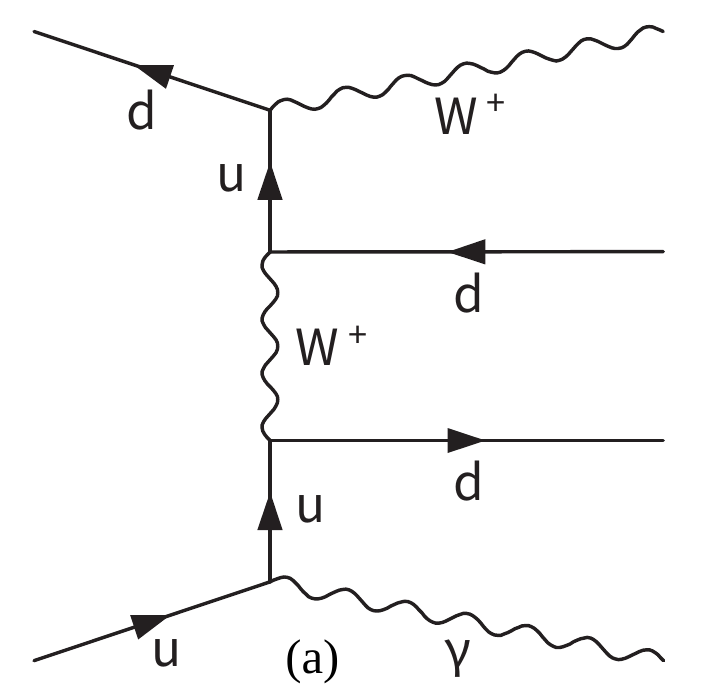}
    \includegraphics[height=0.21\textwidth,width=0.23\textwidth]{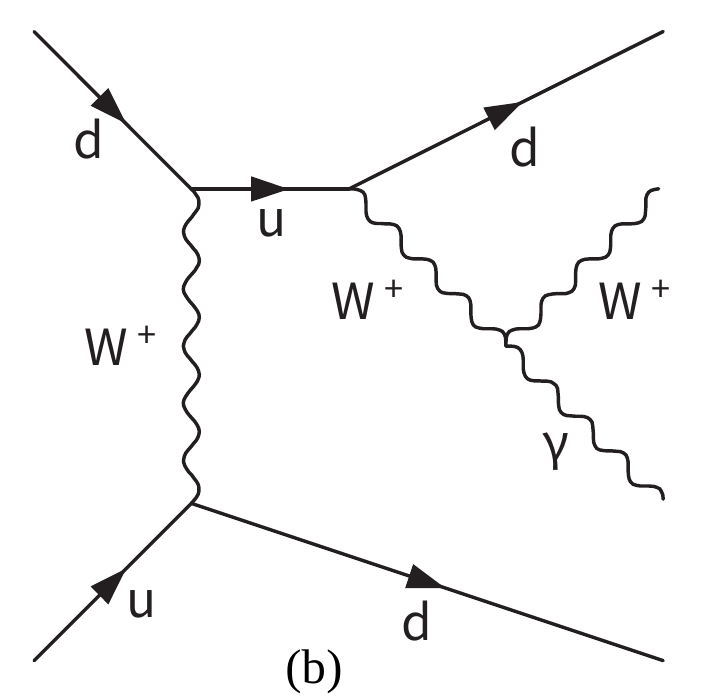}
    \includegraphics[height=0.21\textwidth,width=0.23\textwidth]{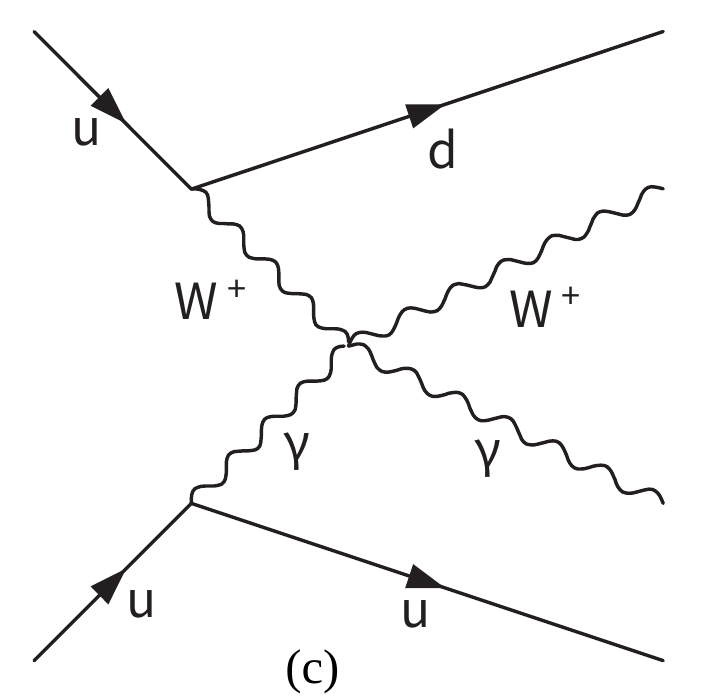}
    \includegraphics[height=0.21\textwidth,width=0.23\textwidth]{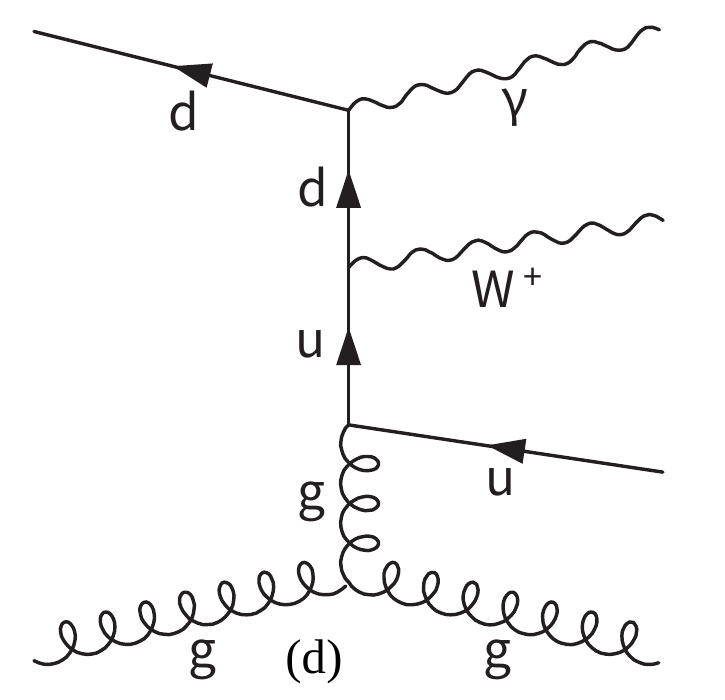}
  \caption{Representative Feynman diagrams for \Wgjj production at the LHC: EW (a), EW through triple (b) and quartic (c) gauge boson couplings, and QCD-induced (d).}
  \label{fig:wa_feynman}
\end{figure*}

The EW signal includes both VBS and non-VBS diagrams,
such as the contributions depicted in Figs.~\ref{fig:wa_feynman}(a)-\ref{fig:wa_feynman}(c).
The QCD-induced production of \Wgjj, in which both jets originate from QCD interaction,
occurs at a much higher rate and is depicted in the rightmost diagram in Fig.~\ref{fig:wa_feynman}(d).
The interference among the VBS diagrams ensures the unitarity of the VBS cross section in the SM at high energy. An interference is also expected between the EW- and QCD-induced processes~\cite{Khoze:2002fa,PhysRevD.69.093004}. The interference is regarded as a background when measuring the EW process.
The cross section for the EW \Wgjj production and the total cross section for the \Wgjj production that includes both the EW- and QCD-induced processes are determined in the same restricted fiducial region. The measurements are based on a two-dimensional fit in the invariant mass $m_{\ell\gamma}$ of the lepton and the photon and the invariant mass $m_{\text{jj}}$ of the two jets.
Differential cross sections unfolded to the particle level are also measured.

In addition, BSM couplings, such as anomalous triple and quartic gauge couplings (aTGCs and aQGCs), as predicted in BSM theories~\cite{paper_aqgc} and would affect the \Wgjj production.
The aTGCs are well constrained by processes such as Higgs boson and diboson production, 
whereas the aQGCs can be better constrained by VBS measurements.
In this analysis, constraints are placed on aQGCs in terms of
dimension-8 effective field theory operators.

The data set used in this analysis corresponds to an integrated luminosity of 138\fbinv
collected in Run 2 with the CMS detector~\cite{Chatrchyan:2008zzk} at the LHC. 
The final state is characterized by an isolated electron or muon with high transverse momentum (\pt), large missing transverse momentum (\ptmiss) from the leptonic decay of the \PW boson, a high-\pt~isolated photon, and two jets.
Exploiting the VBS \Wgjj topology,  the two jets are required to have a large invariant mass \mjj and a large separation in pseudorapidity $\abs{\dejj}$.
This selection effectively suppresses the contamination from the QCD-induced production of \Wgjj,
as well as the non-VBS EW contribution [Figs.~\ref{fig:wa_feynman}(a) and ~\ref{fig:wa_feynman}(b)].

\section{The CMS detector}
\label{sec:detector}
The central feature of the CMS~\cite{Chatrchyan:2008zzk} apparatus is
a superconducting solenoid of 6\unit{m} internal diameter, providing a magnetic field of 3.8\unit{T}.
Within the solenoid volume are a silicon pixel and strip tracker, a lead tungstate crystal electromagnetic calorimeter (ECAL), and a brass and scintillator hadron calorimeter (HCAL), each composed of a barrel and two endcap sections.
Forward calorimeters extend the coverage provided by the barrel and endcap detectors up to a pseudorapidity of $\abs{\eta}=5$.
Muons are detected in gas-ionization chambers embedded in the steel flux-return yoke outside the solenoid.
Events of interest are selected using a two-tiered trigger system~\cite{Khachatryan:2016bia,CMS:2020cmk}.
The first level, composed of specialized hardware processors,
uses information from the calorimeters and muon detectors to select events at a rate of around 100\unit{kHz} within a fixed latency of about 4\mus.
The second level, the high-level trigger (HLT), consists of a farm of processors running a version
of the full event reconstruction software optimized for fast processing that reduces the event rate to around 1\unit{kHz} before data storage.
A more detailed description of the CMS detector, together with a definition
of the coordinate system and kinematic variables, is reported in Ref.~\cite{Chatrchyan:2008zzk}.

\section{Signal and background simulation}
\label{sec:simulation}
The signal and background processes are simulated using the \MGvATNLO (MG5) Monte Carlo (MC) generator~\cite{MGatNLO}.
The EW \Wgjj signal is simulated at leading order (LO) using MG5 version 2.6.0.
The dominant background from the QCD-induced production of \Wgjj is simulated with up to one additional jet in the matrix element calculations at next-to-leading order (NLO) with MG5 version 2.4.2, using the FxFx scheme~\cite{Frederix:2012ps} to merge jets from matrix elements and from parton showering.
The interference term between the EW- and QCD-induced processes, of order $\mathcal{O}(\alpha^4\alpS)$ at tree level, is estimated with a full simulation
and is treated as a part of the QCD-induced \Wgjj contribution.
 The contribution of the interference is calculated as the difference
between the total \Wgjj production, which contains the interference term, 
and the sum of the individual EW- and QCD-induced \Wgjj contributions as simulated by MG5.
The interference term ranges from 1\% to 3\% of the expected EW signal in the signal region (defined in Section~\ref{sec:selection}), varying with \mjj bin.

Other background contributions include diboson processes (\VV = \WW, \WZ, \ZZ) simulated
at LO with \PYTHIA~8.212~\cite{Sj_strand_2015},
top quark processes (\ttbar and single top) simulated at NLO
with \POWHEG~2.0~\cite{Nason:2004rx,Frixione:2007vw,Frederix:2012dh,Alioli:2009je,Re:2010bp},
\ttg production simulated at NLO with MG5 using the FxFx jet merging scheme,
and \Zg simulated at NLO with MG5.

The \PYTHIA 8 generator with the CUETP8M1~\cite{Skands:2014pea,Khachatryan:2015pea} tune for 2016
and the CP5~\cite{Sirunyan:2019dfx} tune for 2017--2018
is used for parton showering, hadronization, and underlying-event simulation.
The NNPDF~3.0 (3.1) set~\cite{Ball:2014uwa} is used for the parton distribution functions (PDFs)
for the simulated samples of the 2016 (2017-2018) data-taking periods.
All simulated events are processed with \GEANTfour~\cite{geant4} for the CMS detector simulation.
Correction factors evaluated with the tag-and-probe method~\cite{tagandprobe} are used to
account for differences between data and simulation in the trigger, reconstruction, and identification (ID) efficiencies.
Additional simulated \pp interactions (pileup, PU) are superimposed over the hard scattering interaction
with a distribution matching that obtained from the collision data.

\section{Object reconstruction}
\label{sec:reconstruction}
The particle-flow (PF) algorithm~\cite{Sirunyan:2017ulk} reconstructs
and identifies individual particles in an event, through an optimized
combination of information from the various components of the CMS detector.
The energy of photons is obtained from the ECAL measurement.
The energy of electrons is determined from a combination of
the electron momentum at the primary interaction vertex from the tracker,
the energy of the corresponding ECAL cluster, and the energy sum of
all bremsstrahlung photons spatially compatible with originating from the electron track.
The energy of muons is obtained from the curvature of the corresponding tracks.
The energy of charged hadrons is determined from a combination of their momentum
measured in the tracker and the matching ECAL and HCAL energy depositions,
corrected for the response of the calorimeters to hadronic showers.
The energy of neutral hadrons is obtained from the corresponding corrected ECAL and HCAL energies.
The PF candidates are used for a variety of purposes in this analysis,
such as evaluating electron, muon, and photon isolation variables,
reconstructing jets, and computing the \ptmiss in the event, as described below.

The reconstructed vertex with the largest value of summed physics-object $\pt^2$ is taken as the primary \pp interaction vertex~\cite{TRK-11-001}.
The jets are clustered using the anti-\kt jet finding algorithm~\cite{Cacciari:antikt,Cacciari:fastjet1}
using tracks assigned to candidate vertices as inputs and the distance parameter is set to 0.4.

Electron candidates must satisfy $\abs{\eta}<2.5$ and $\pt>35\GeV$,
excluding the ECAL transition region $1.444<\abs{\eta}<1.566$.
Electrons are also required to satisfy identification criteria~\cite{Khachatryan:2015hwa}:
a selection on the relative amount of energy deposited in the HCAL,
a match of the trajectory in the tracker with the position of the ECAL cluster,
requirements on the number of missing measurements in the tracker,
the compatibility of the electron track and the primary vertex,
and \sigmaee, which quantifies the spread along $\eta$ of the shower in the ECAL.
Electrons identified as arising from photon conversions are removed~\cite{Khachatryan:2015hwa,electron_7tev}.
The CMS cut-based tight ID is used to define tight electrons from \PW decays,
whereas the CMS cut-based veto ID is used to define loose electrons to suppress events that contain additional leptons.
An isolation requirement is applied to electrons.
The isolation variable is defined relative to the electron \pt by summing the \pt of charged hadrons
and neutral particles within geometrical cones of $\DR=\sqrt{\smash[b]{(\Delta\eta)^2+(\Delta\phi)^2}}=0.3$
around the electron momentum direction.
To minimize PU effects, only charged hadrons originating from the primary vertex are included.
For the neutral-hadron and photon components, an estimate of the expected PU contribution (\PUpt) is subtracted~\cite{Sirunyan_2020}.
For the tight (loose) electrons, the isolation variable is required to be less than $0.0287 + 0.506\GeV / \pt~(0.198 + 0.506\GeV / \pt)$
if the pseudorapidity of the ECAL cluster (\etaSC) satisfies $\abs{\etaSC}<1.479$,
and less than $0.0445 + 0.963\GeV / \pt~(0.203 + 0.963\GeV / \pt)$ if $1.479<\abs{\etaSC}<2.5$.

Muon candidates are required to satisfy $\abs{\eta}<2.4$ and $\pt>35\GeV$.
They must satisfy ID criteria based on the number of measurements in the muon system and the tracker,
the number of matched muon detector planes, the quality of the combined fit to the track,
and the compatibility of the muon to originate from the primary vertex~\cite{Sirunyan_2018}.
The CMS cut-based tight ID is used.
An isolation requirement is applied to muons.
The isolation variable is defined relative to the muon \pt by summing the \pt of charged hadrons
and neutral particles within geometrical cones of $\DR=0.4$.
The PU suppression is performed in a similar way as done for electrons.
The isolation variable is required to be $<0.15 (0.25)$ to define tight (loose) muons.
Tight muons are used to select signal events,
whereas loose muons are used to veto events that feature additional leptons~\cite{Sirunyan_2018}.

Photon candidates must satisfy $\abs{\eta}<2.5$ and  $\pt>25\GeV$,
excluding the ECAL transition region of $1.444<\abs{\eta}<1.566$.
To minimize the contribution of jets misidentified as photons, photon candidates must satisfy~\cite{photon_8tev}
criteria based on the distribution of energy deposited in the ECAL and HCAL, and
criteria based on the isolation variables constructed from the kinematic inputs of the charged hadrons, neutral hadrons, and other photons near the photon of interest.
The CMS cut-based medium ID defines tight photons and is used to identify prompt photons
(\ie, not originating from hadron decays) in the final state,
and the CMS cut-based loose ID defines loose photons and is used to 
identify nonprompt photons, which are mainly products of neutral pion decays~\cite{photon_8tev}.
An isolation requirement using a consistent definition as mentioned above for electrons and muons 
is applied with $\DR=0.3$ for the three components separately, \ie, the charged hadron isolation must be less than 1.141 (1.051),
the neutral hadron isolation must be less than
$1.189 + 0.01512\pt + 2.259\times 10^{-5} \pt^2$
($2.718 + 0.0117\pt + 2.3\times 10^{-5} \pt^2$)
and the photon isolation component must be less than
$2.08 + 0.004017 \pt$ ($3.867 + 0.0037 \pt$),
for the tight photon candidates found in the barrel (endcap) region,
whereas the charged hadron isolation must be less than 1.694 (2.089),
the neutral hadron isolation must be less than
$24.032 + 0.01512\pt + 2.259\times 10^{-5} \pt^2$
($19.722 + 0.0117\pt + 2.3\times 10^{-5} \pt^2$)
and the photon isolation component must be less than
$2.876 + 0.004017 \pt$ ($4.162 + 0.0037 \pt$),
for the loose photon candidates found in the barrel (endcap) region, where \pt is measured in \GeV.
The PU suppression is performed in a similar way as for electrons.
An additional veto is applied on electrons reconstructed as photons.

Jets are required to have $\abs{\eta}<4.7$ and $\pt>50\GeV$.
To reduce the contamination from PU, charged PF candidates within the tracker acceptance are excluded from the jet clustering when they are associated with PU vertices~\cite{Sirunyan:2017ulk}.
The contribution from neutral PU particles to the jet energy is corrected based on
the projected area of the jet onto the front face of the calorimeter~\cite{jetarea_method}.
A jet energy correction, similar to the one developed for 8\TeV collisions~\cite{jer},
is obtained from dedicated studies performed on both data and simulated events
(typically involving dijet, {\Pgg}+jet, {\PZ}+jet, and multijet production).
Other residual corrections are applied to the data as functions of \pt and $\eta$
to correct for small differences between data and simulation.
Additional quality criteria are applied to jet candidates to remove spurious jet-like features
originating from isolated noise patterns in the calorimeters or in the tracker~\cite{CMS:2010xta}.

The missing transverse momentum \ptvecmiss is computed as the projection onto the plane perpendicular to the beam axis of the negative vector momentum sum of all PF candidates originating from the primary vertex in an event~\cite{Sirunyan:2019kia}, and its magnitude is denoted as \ptmiss.
The jet energy corrections are propagated to the \ptvecmiss.
Data-to-simulation efficiency ratios are used as scale factors to correct the simulated event yields.

\section{Event selection}
\label{sec:selection}
Signal event candidates are collected with single lepton triggers and are selected by requiring exactly one electron (muon)
with $\pt>35\GeV$ and \mTW $>30\GeV$, where
\mTW is the transverse mass of the \PW boson defined as $\sqrt{\smash[b]{2\ptlep\ptmiss[1-\cos{(\dphilmet)}]}}$, \ptlep is the lepton \pt, and \dphilmet is the azimuthal angle between the \ptlep and the \ptvecmiss directions.
Events are required to contain a well-identified and isolated photon with $\ptg>25\GeV$,
$\ptmiss>30\GeV$, and at least two jets, each with $\abs{\eta}<4.7$ and $\pt>50\GeV$.
A separation of $\DR>0.5$ is required between any two selected objects (photon, lepton, jets), as detailed in Section~\ref{sec:fidewxsmeas}.
In the electron channel, we additionally require the invariant mass $\mlepg$
of the selected photon and electron to be inconsistent with the \PZ boson mass,
$\abs{\mlepg-m_{\PZ}}>10\GeV$, to suppress the $\PZ\to\Pep\Pem$ background where one electron is misidentified as a photon.
Depending on the photon pseudorapidity, the electron and muon channels
are each subdivided into a barrel region with $\abs{\etag}<1.444$, and an endcap region with $1.566<\abs{\etag}<2.5$.
The nominal selection consists of all the above requirements.

The longitudinal component of the neutrino momentum is estimated by solving
the quadratic equation that constrains the mass of the charged lepton and neutrino system
to the world-average value of the \PW boson mass~\cite{PDG2018}.
As described in Ref.~\cite{Sirunyan2018}, when there are multiple solutions,
the one with the smallest longitudinal neutrino momentum component is chosen;
if there are only complex solutions, the real part is chosen as the longitudinal momentum.

The signal region (SR) is defined as the above nominal selection with the additional requirements of
$\mjj>500\GeV$, $\abs{\dejj}>2.5$, $\mWg>100\GeV$, $\kinsysY<1.2$~\cite{zeppenfeld},
and $\kinsysPhi>2$, where \mWg, \phiWg, and $y_{\Wg}$ are the invariant mass, azimuthal angle, and the rapidity of the \Wg system, respectively, \phiDijet is the azimuthal angle of
the dijet system between the two \pt-leading jets, and $y_{j1(2)}$ is the rapidity of the \pt-leading (subleading) jet.
The requirements on \kinsysY and on \kinsysPhi are intended to
ensure that the momentum of the \Wg system is balanced by that of the dijet system,
which is expected in the absence of additional QCD radiation.
The selection thresholds are determined by scanning the expected significance of the EW signal to give the maximum sensitivity.

A control region (CR) is defined to validate the modeling from simulation and perform a background estimation derived from data.
The CR uses the nominal selection mentioned above with the additional requirements of $200<\mjj<500\GeV$.
The contamination from signal events in the CR is less than 1\%.

\section{Background estimation}
\label{sec:bkg_estimation}

\begin{figure}[htb!]
  \centering
  \includegraphics[width=0.47\textwidth]{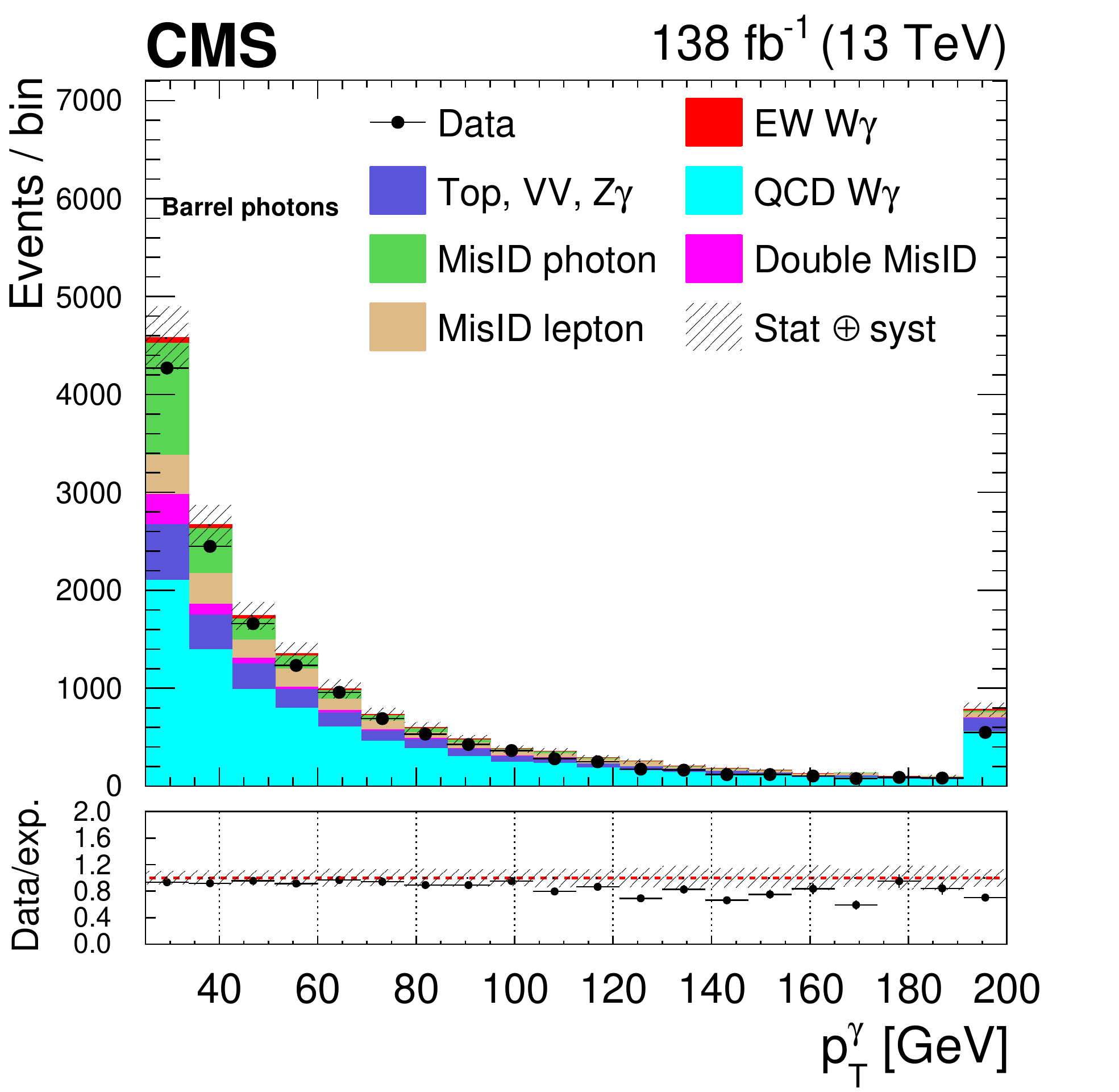}
  \includegraphics[width=0.47\textwidth]{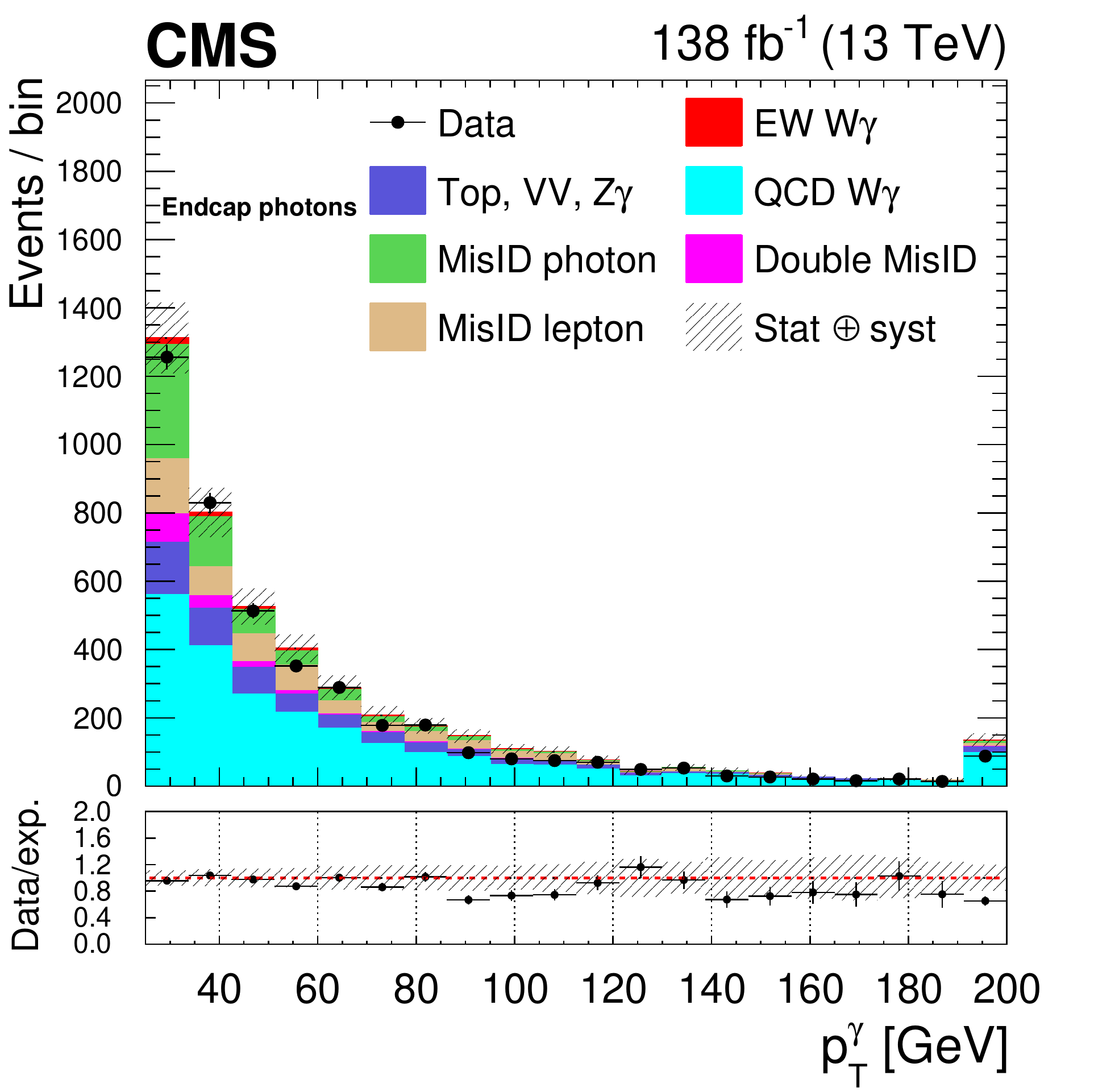}
  \caption{The \pt distributions for photons in the barrel (\cmsLeft) and in the endcaps (\cmsRight) in the control region for data and from background estimations before the fit to the data. The misID backgrounds are derived from data, whereas the remaining backgrounds are estimated from simulation. All events with a photon $\pt>200\GeV$ are included in the last bin. The hatched bands represent the combined statistical and systematic uncertainties on the predicted yields. The vertical bars on the data points represent the statistical uncertainties of data. The bottom panels show the ratios of the data to the predicted yields.}
    \label{fig:control_leptpt}
\end{figure}

In Fig.~\ref{fig:control_leptpt} the $\ptX{\gamma}$ distributions for the unfit data and the estimated backgrounds in the CR are presented for the barrel (\cmsLeft) and endcap (\cmsRight).
This region is used to constrain the QCD \Wgjj background.
The estimations of the backgrounds are described in this section.

Reconstructed photons or leptons that do not originate from the hard interaction are denoted as misidentified (misID) photons and leptons.
This reducible background includes genuine photons or leptons, as well as photons or leptons of instrumental origin.
Because of the variety of sources of these misID particles and the difficulty of modeling instrumental effects, their contribution is estimated using data in a signal free region.

The main backgrounds arise from \Wpjets and top quark processes where the jet constituents are misidentified as a photon.
The method used to estimate this background involves measuring the fraction of jets misidentified as photons in data and applying a per-photon extrapolation factor from the region with loose photons to the signal region with tight photons. 
The factors are extracted as functions of the photon \pt and $\eta$.
The fraction of jets misidentified as photons is determined from a template fit to the photon \sigmaee observable, 
which is the lateral extension of the shower,
defined as the energy-weighted spread within the 5$\times$5 crystal matrix centered on the
crystal with the largest energy deposit in the supercluster.
The prompt photons are more populated in the small \sigmaee region, while the nonprompt photons are enriched in the large \sigmaee region.
The fit template for the prompt photons uses MC, while the fit template for the nonprompt photons uses data from a sideband of the photon isolation distribution in \Wpjets using the same method as used in Ref.~\cite{Sirunyan:2020tlu}.

The background from jets misidentified as leptons (nonprompt leptons) is estimated in a similar way.
The lepton misidentification rate \frlep is defined as the ratio of
the number of misID leptons passing the tight lepton requirements to the number of leptons passing only the loose lepton requirements.
To extrapolate from loose to tight requirements leptons, an extrapolation factor is defined as:
${\frlep}/({1 - \frlep})$.
To suppress additional contamination from genuine leptons,
the \Wpjets and \Zpjets contributions are subtracted from both the numerator and denominator using MC simulation.
The extrapolation factor is measured as a function of the $\eta$ and \pt of the lepton in a CR dominated by dijet events.
This CR is defined by selecting one lepton, one jet well separated from the lepton, and $\ptmiss < 30\GeV$.
More details are described in Ref.~\cite{Sirunyan:2017ret}.

The double-misID background is defined as events containing both a misID photon and a misID lepton.
Its yield is estimated using an event sample where both the photon and lepton are required to pass the loose lepton requirements and fail the tight lepton requirements.
A weight is assigned to such events, equal to the product of the misID extrapolation factors of the photon and lepton.
Double-misID events contaminate the single-misID background estimate since the second object is assumed to be genuine.
Whenever a weight is added to the double-misID estimate, the same weight is subtracted from both the single-photon and -lepton estimates.
In addition, events in which genuine photons and leptons pass the loose lepton requirements
but fail the tight lepton requirements contaminate both the single- and double- misID estimates.
This source of contamination is estimated and removed using simulation with reconstructed objects matched to generator-level objects.

Other background contributions that feature genuine photons and leptons in the final state,
such as top quark, diboson and $\PZ\gamma$,
are estimated from MC simulation and normalized to the integrated luminosity
of the data set using their corresponding cross sections.

\section{Systematic uncertainties}
\label{sec:systematics}
Systematic uncertainties that affect the measurements arising from experimental inputs,
such as detector effects and methods, and theoretical inputs such as the choice of the renormalization
(\muR) and factorization (\muF) scales and the choice of PDF sets, are included.
Each source of systematic uncertainty is quantified by evaluating its effect
on the yield and on the distributions of relevant kinematic variables in the signal and background categories.
The uncertainties are calculated bin-by-bin and propagated to the final distributions.

The uncertainties in jet energy scale (JES) and jet energy resolution (JER) are estimated by shifting or spreading the jet energies in the simulations up and down by one standard deviation,
and are then propagated to all relevant variables, including VBS jet kinematic observables and \ptmiss,
and the impacts on the signal and background yields are evaluated.
The uncertainties arising from the JES and JER correspond to various processes and
various \mjj--\mlepg (\mjj vs. \mlepg 2D distribution) bins are in the ranges of 0.1--34\% and 1.8--33\%, respectively.
The uncertainties in the lepton trigger, reconstruction, and selection efficiencies,
measured using a tag-and-probe technique, are 1.8--4.6\%~\cite{Sirunyan_2018,Khachatryan:2015hwa}.
The uncertainties in the photon reconstruction and selection efficiencies
are 1.9--4.3\%~\cite{CMS:2020uim}.
The integrated luminosities have uncertainties in the 1.2--2.5\% range~\cite{CMS-LUM-17-003,CMS-PAS-LUM-17-004,CMS-PAS-LUM-18-002}, with an overall uncertainty for the 2016--2018 dataset of 1.6\%.

The statistical uncertainties arising from the limited size of both the simulated
and data samples used in our background and signal predictions are estimated assuming a Poisson distribution.
The uncertainties related to the limited number of simulated events or
to the limited number of events in the data control samples are 1.2--11\%
for the EW \Wgjj signal, 2.1--48\% for the QCD-induced \Wgjj background, 4.9--77\%
for the nonprompt-lepton background, and 2.1--45\% for the nonprompt-photon background.
Some of these statistical uncertainties increase with increasing \mjj and \mlepg. The largest values typically come from bins where the specific process is less important, and do not significantly impact the signal sensitivity.
All the statistical uncertainties are uncorrelated across various processes and
bins of any single distribution.

An overall systematic uncertainty in the nonprompt-photon background estimate
is defined as the quadratic sum of the systematic uncertainties from three distinct sources.
The uncertainty arising from the choice of the isolation variable sideband is
evaluated by estimating the nonprompt-photon fraction with alternative choices of the sideband~\cite{Khachatryan:2017jub}. The statistical uncertainty in extracting the fake photon fraction is obtained from the template fits.
The nonclosure uncertainty is defined by performing the nonprompt-photon fraction fits
using simulated events and comparing the results with the predicted fractions from MC simulation.
The nonclosure uncertainty in the endcap region is larger than
in the barrel region and increases with the photon \pt.
The overall systematic uncertainty in the nonprompt-photon background
ranges from 7.8\% to 12\%, dominated by the nonclosure contribution.

Similarly, the uncertainty in the nonprompt-lepton estimate
comes from the nonclosure that is obtained using MC samples.
The same misidentified lepton method used in the analysis is applied to MC \gpjets events,
and the result is compared with the true number of \gpjets events falling into the SR.
The difference of the two quantifies the nonclosure.
The selection used is the same as in the nominal event selection, 
except that the \mTW and \ptmiss requirements are removed to increase the size of the selected sample.
The uncertainty associated with the nonprompt-lepton background is 30\%.

The effects of the choice of \muR and \muF in the theoretical calculation
for signal and background cross sections are estimated by independently changing
\muR and \muF up and down by a factor of 2 from their nominal values in each event, satisfying $1/2<\muR/\muF<2$.
The uncertainties are defined as the maximal differences from the nominal values.
The PDF uncertainties are evaluated according to the procedure
described in Ref.~\cite{nnpdf_4lhc} using the NNPDF set.
For the signal, the scale uncertainty varies within 0.7--5.4\% and the PDF uncertainty varies within 0.06--0.10\% in the acceptance.
The scale uncertainty in the QCD-induced \Wgjj process corresponds to a 0.08--12\% uncertainty in the acceptance.
It is constrained by the simultaneous fit to the data in the CR.
The PDF uncertainty in the acceptance of the QCD-induced \Wg production is 0.05--1.40\%.

A correction factor is applied to the simulated events to account
for the first level trigger timing drift in 2016 and 2017 data~\cite{CMS:2020cmk}.
This mistiming results in a loss of trigger efficiency in the data and is not modeled by the simulation.
Uncertainties arising from these correction factors vary within 0.9--3.4\%,
and are treated as correlated across various processes and bins of the 2016 and 2017 data analysis.

\section{Observation of EW \texorpdfstring{\Wg}{W gamma} production}
\label{sec:ewsig}

The measurement of the total EW \texorpdfstring{\Wg}{W gamma} production rate
is performed using a binned likelihood fit to the data of the two-dimensional (2D) distribution in \mjj (four bins) and \mlepg (three bins).
Both \mjj and \mlepg are highly discriminating variables between the EW signal and  the QCD-induced \Wgjj background. Furthermore, the 2D analysis provides a larger expected significance than using either variable alone.

\begin{figure*}[htb!]
  \centering
      \includegraphics[width=0.47\textwidth]{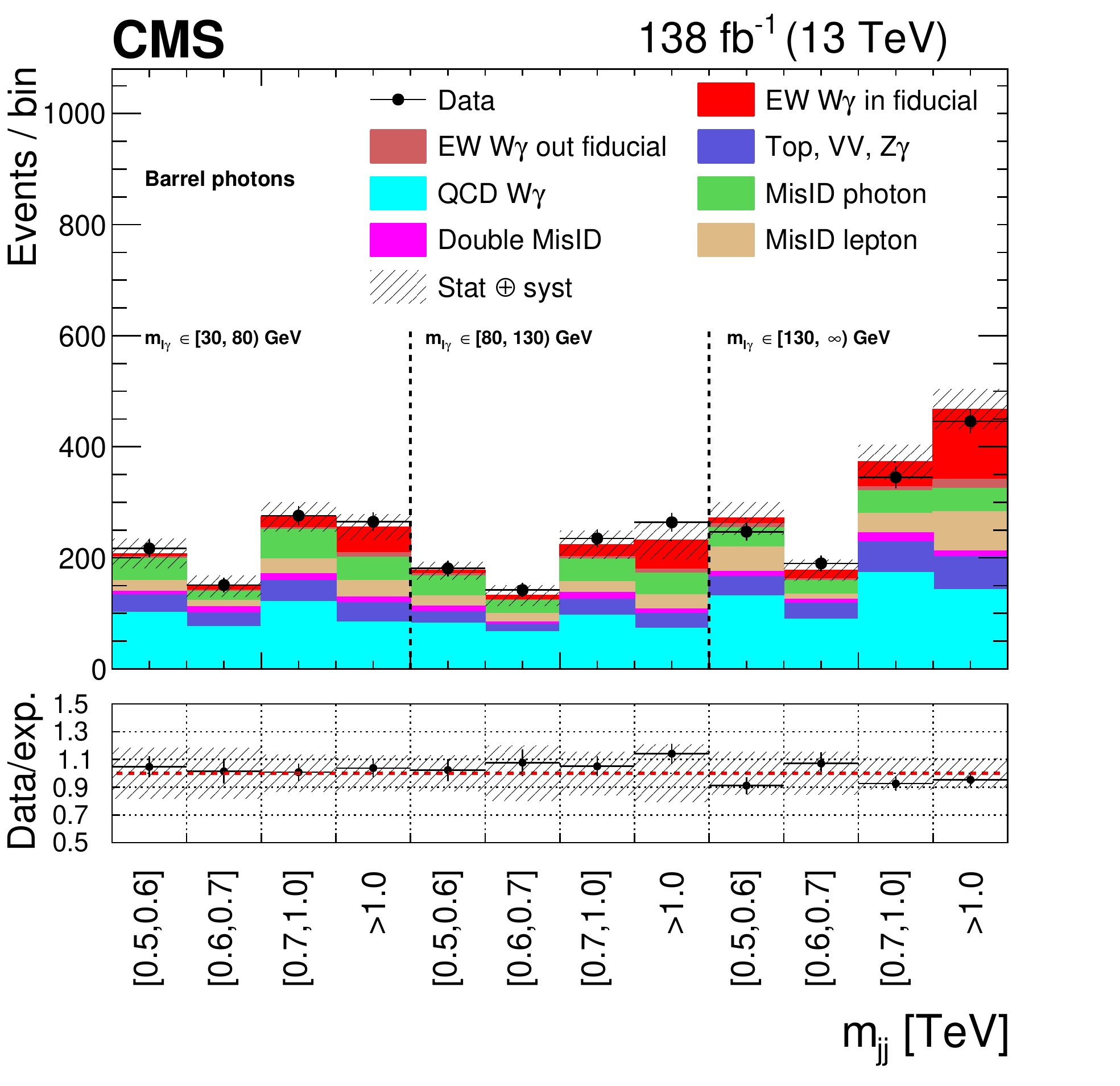}
      \includegraphics[width=0.47\textwidth]{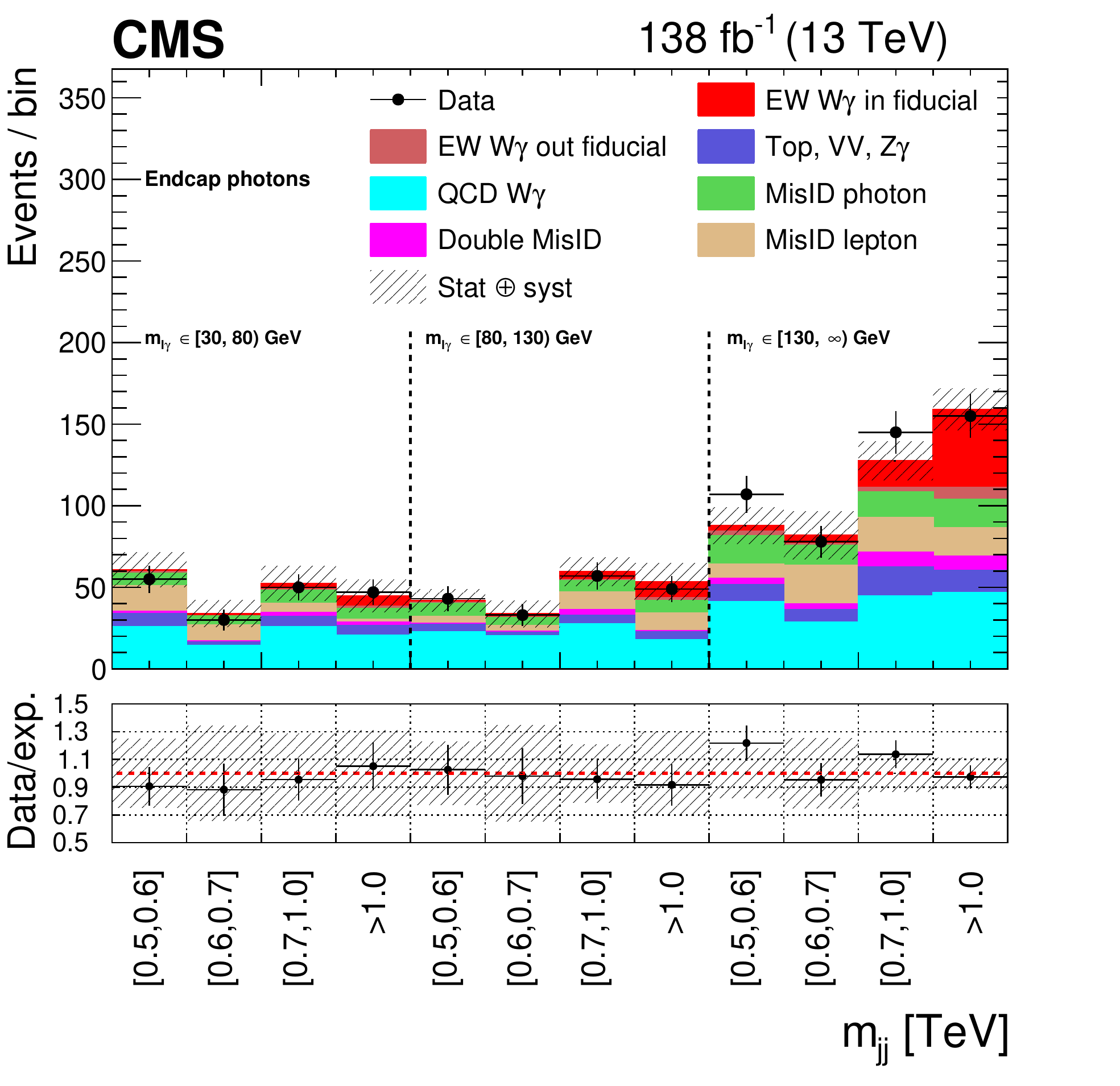}
      \caption{The 2D distributions used in the fit for the total EW \texorpdfstring{\Wg}{W gamma} cross section measurement.
		The hatched bands represent the combined statistical and systematic uncertainties in the predicted yields.
                The vertical bars on the data points represent the statistical uncertainties of data.
		The expectation is shown after the fit to the data.
                EW \Wg inside (outside) fiducial region stands for the events of EW \Wg falling inside (outside) the fiducial region defined in Section~\ref{sec:fidewxsmeas}.
		}
      \label{fig:2d_strength}
\end{figure*}

Data in the SR and CR are both included in the fits to constrain the dominant background (QCD-induced \Wgjj).
Table~\ref{tab:yields} shows the signal and background yields after the fit, as well as the observed data yields.
Figure~\ref{fig:2d_strength} shows the observed and expected distributions of \mjj--\mlepg used in the total EW \Wgjj cross section measurement. 
The expectation is given after the fit to data.

\begin{table*}[htbp]
\centering
  \topcaption{Number of \Wg events from the fit to the data in the signal region. The signal predictions inside and outside the fiducial region defined in Section~\ref{sec:fidewxsmeas} are shown. The contributions of various backgrounds are also shown. Statistical and systematic uncertainties are added in quadrature.}
\begin{scotch*}{lcc}
                        &  Barrel   &  Endcap       \\
\hline
EW \Wg inside fiducial region &  316 $\pm $ 16 &  90.2 $\pm $ 5.5    \\
EW \Wg outside fiducial region &  64.7 $\pm $ 2.0 &  20.4 $\pm $ 1.0 \\
QCD \Wg &  1301 $\pm $ 28 &  362 $\pm $ 13 \\
top, VV,~\Zg &  402 $\pm $ 14 &  93.3 $\pm $ 7.2 \\
Nonprompt photon &  434 $\pm $ 13 &  120.2 $\pm $ 5.7 \\
Nonprompt muon &  134 $\pm $ 27 &  45 $\pm $ 11 \\
Nonprompt electron &  189 $\pm $ 20 &  86 $\pm $ 13  \\
Nonprompt photon, nonprompt muon &  43.0 $\pm $ 7.0 &  14.6 $\pm $ 3.4 \\
Nonprompt photon, nonprompt electron &  75.5 $\pm $ 5.5 &  25.0 $\pm $ 2.0\\
\hline
Total prediction &  2960 $\pm $ 43 &  856 $\pm $ 21\\
Data &  2959 $\pm $ 57 &  849 $\pm $ 32 \\
\end{scotch*}
\label{tab:yields}
\end{table*}

The signal significance is quantified using a profile likelihood test statistic~\cite{Wilks:1938dza}.
This test statistic involves the ratio of two Poisson likelihood functions,
one in which the signal strength is fixed to zero and
one in which the signal strength is allowed to have any positive value.
The signal strength represents the ratio of observed to expected signal yields.
Systematic uncertainties are included as nuisance parameters in the likelihood function
that scale the relevant processes using log-normal probability density functions.
The distribution of the test statistic is assumed to be in the asymptotic regime
where there is a simple relationship between its value and the significance of the result~\cite{CLs}.
The observed (expected) significance is 6.0 (6.8) SD for the EW \Wg processes. %6.03 (6.79) SD.

\section{Fiducial cross section measurement}
\label{sec:fidewxsmeas}

The fiducial cross section measurement for the EW \Wg production
at 13\TeV is extracted with the same 2D \mjj--\mlepg binning used for the signal significance.
The fiducial region is defined based on the particle-level (for leptons, photons, jets) quantities:
one lepton $\ptlep>35\GeV$ and $\abs{\etalep}<2.4$, $\ptmiss>30\GeV$,
$\ptg>25\GeV$, $\abs{\etag}<1.444$ or $1.566<\abs{\etag}<2.5$, $\DR_{\Pell\gamma}>0.5$, $\mTW>30\GeV$,
and two jets with $\ptX{j1(2)}>50\GeV$, $\abs{\etaj}<4.7$, $\mjj>500\GeV$, $\DR_{\jj}>0.5$,
$\DR_{\text{j}\Pell}>0.5$, $\DR_{\text{j}\gamma}>0.5$, and $\abs{\dejj} > 2.5$.
The leptons are reconstructed at the particle level with fully recovered final-state radiation.
The acceptance is defined as the fraction of the signal events passing the fiducial region selection, and is estimated using MG5.
The theoretical uncertainty in the extrapolation between the fiducial and SR is negligible ($<1\%$).
We define the cross section as $\sigfid = \sigmag\,\muhat\,\alphagf$,
where the cross section for the signal events is $\sigmag = 0.776\unit{pb}$ calculated with MG5 at LO in QCD~\cite{MGatNLO},
the observed signal strength parameter $\muhat = 0.88^{+0.19}_{-0.18}$, and the acceptance of the fiducial region, $\alphagf = 0.034$.
The measured fiducial cross section is 
\ifthenelse{\boolean{cms@external}}{
  \begin{multline}
    \sigfidew  = 23.5 \\ \pm 2.8 \stat^{+1.9}_{-1.7}\thy^{+3.5}_{-3.4}\syst\unit{fb} = 23.5^{+4.9}_{-4.7}\unit{fb}.
\end{multline}
}{
\begin{equation}
    \sigfidew = 23.5 \pm 2.8 \stat^{+1.9}_{-1.7}\thy^{+3.5}_{-3.4}\syst\unit{fb} = 23.5^{+4.9}_{-4.7}\unit{fb}.
\end{equation}
}
The observed signal strength is compatible with unity within one SD, and the measured fiducial cross section agrees well with the SM prediction.

The cross section for the sum of the EW and QCD-induced \Wgjj contributions is also measured.
The fiducial region definition is identical to that used for the EW {\Wgjj} fiducial cross section measurement and the formula for the cross section is
$\sigfidewq = \mu (\sigmagew \alphaewgf+\sigmagqcd \alphaqcdgf) \nonumber$.
The inputs used for the fit are similar to the ones for EW {\Wgjj} production, with the difference that EW and QCD-induced {\Wgjj} contributions are combined as signal.
The cross section for QCD-induced {\Wgjj} production is 192.3\unit{pb} calculated with MG5 at NLO in QCD~\cite{MGatNLO},
and \alphaqcdgf is calculated to be $4.6\times 10^{-4}$.
The measured signal strength for the EW+QCD {\Wgjj} production is $0.98^{+0.12}_{-0.11}$
and the observed fiducial cross section is
\ifthenelse{\boolean{cms@external}}{
  \begin{multline}
    \sigfidewq  = 113\\ \pm 2.0 \stat^{+2.5}_{-2.3}\thy^{+13}_{-13}\syst\unit{fb} = 113 \pm 13\unit{fb}.
\end{multline}
}{
\begin{equation}
    \sigfidewq = 113 \pm 2.0 \stat^{+2.5}_{-2.3}\thy^{+13}_{-13}\syst\unit{fb} = 113 \pm 13\unit{fb}.
\end{equation}
}
The observed signal strength is compatible unity within one SD,
the measured fiducial cross section agrees well with the SM prediction.

\section{Differential cross section measurements}

The differential cross sections for the EW only and for the EW+QCD {\Wgjj} production processes
are measured for several characteristic variables
using the same SR as defined in the fiducial cross section measurement.
For each unfolded variable, its generator-level values are mapped to the reconstruction-level ones
in binned histograms that account for the detector resolution effects.
The efficiencies for selecting events from the generator level to the reconstruction level 
are calculated using the same binning as used in the fiducial region measurements,
in order to recover the limited acceptance and selection efficiencies.
Signal events outside the fiducial region are treated as background.
Both the resolution and efficiency effects are evaluated using signal simulation.
A bin-by-bin unfolding is performed to obtain differential distributions,
in which the effects of detector resolution, limited acceptance, and selection efficiencies are corrected.

The unfolded variables include the transverse momentum of the lepton \ptX{\ell}, of the photon \ptX{\gamma};
the invariant masses of the lepton and the photon $m_{\ell\gamma}$;
the transverse momentum of the leading jet (\pt ordered) \ptX{\text{j1}};
the invariant mass of the two jets $m_{\text{jj}}$;
and the separation in pseudorapidity of the two jet $\Delta\eta_{\text{jj}}$.
Since the ranges of some variables extend to infinity,
the last bins accommodate all the events above the last bin boundaries, but the bin widths
that are used in the denominator are finite and are (110, 400), (170 200), (160, 1000), (250, 500), and (1500, 2000) \GeV for \ptX{\ell}, \ptX{\gamma}, $m_{\ell\gamma}$, \ptX{\text{j1}} and $m_{\text{jj}}$, respectively.

\begin{figure*}[htbp!]
   \centering
      \includegraphics[width=0.40\textwidth]{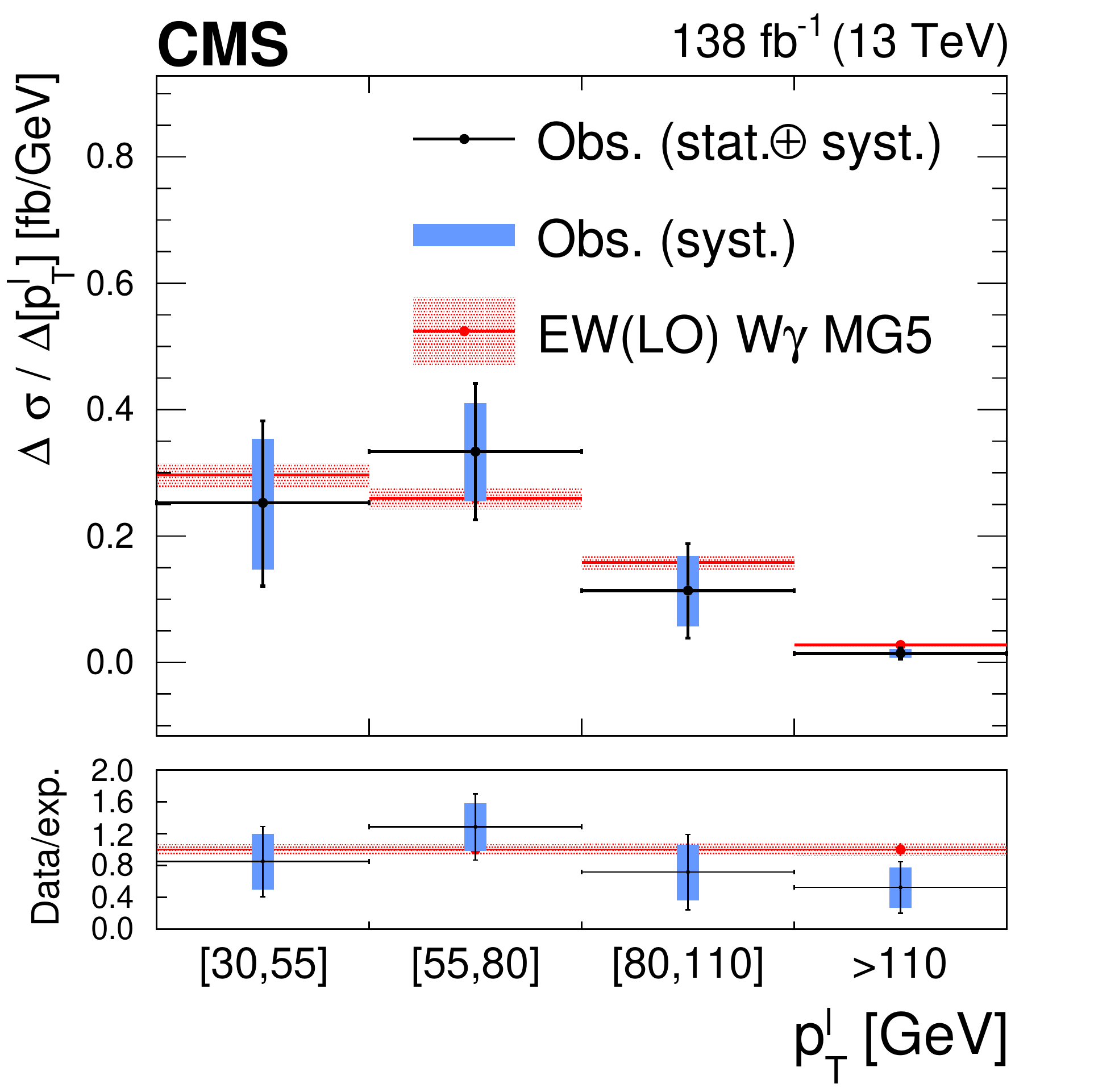}
      \includegraphics[width=0.40\textwidth]{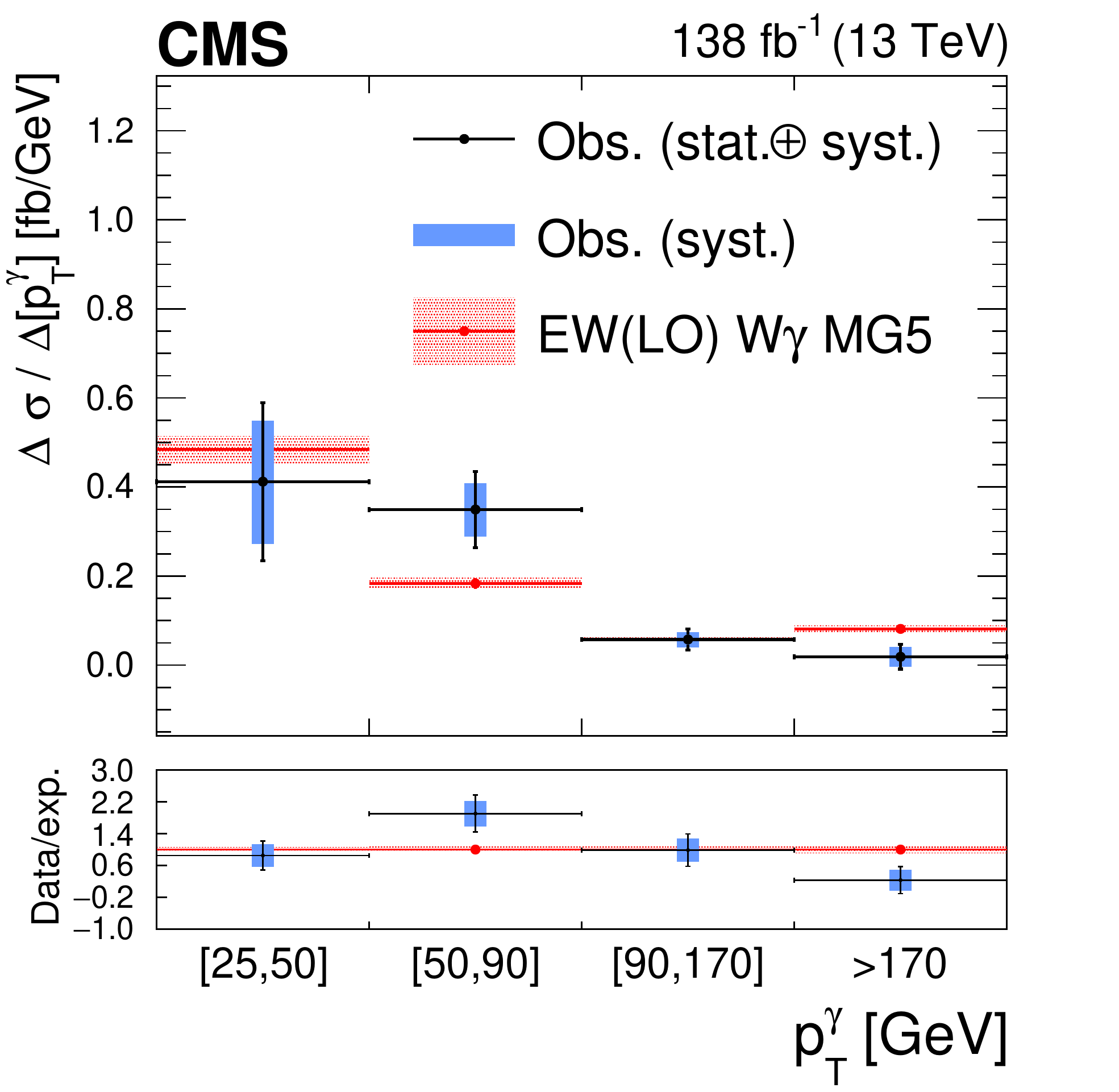}
      \\
      \includegraphics[width=0.40\textwidth]{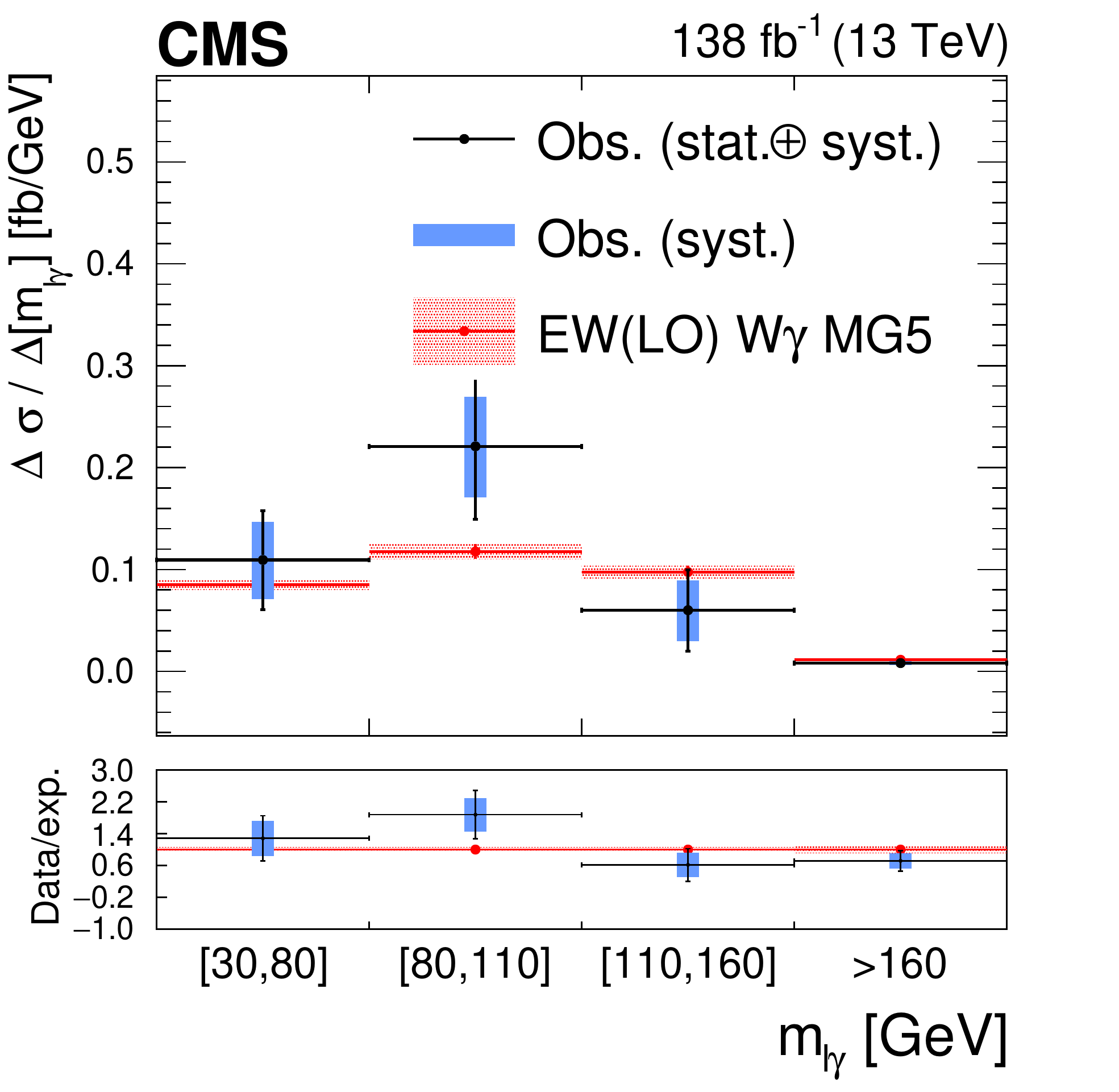}
      \includegraphics[width=0.40\textwidth]{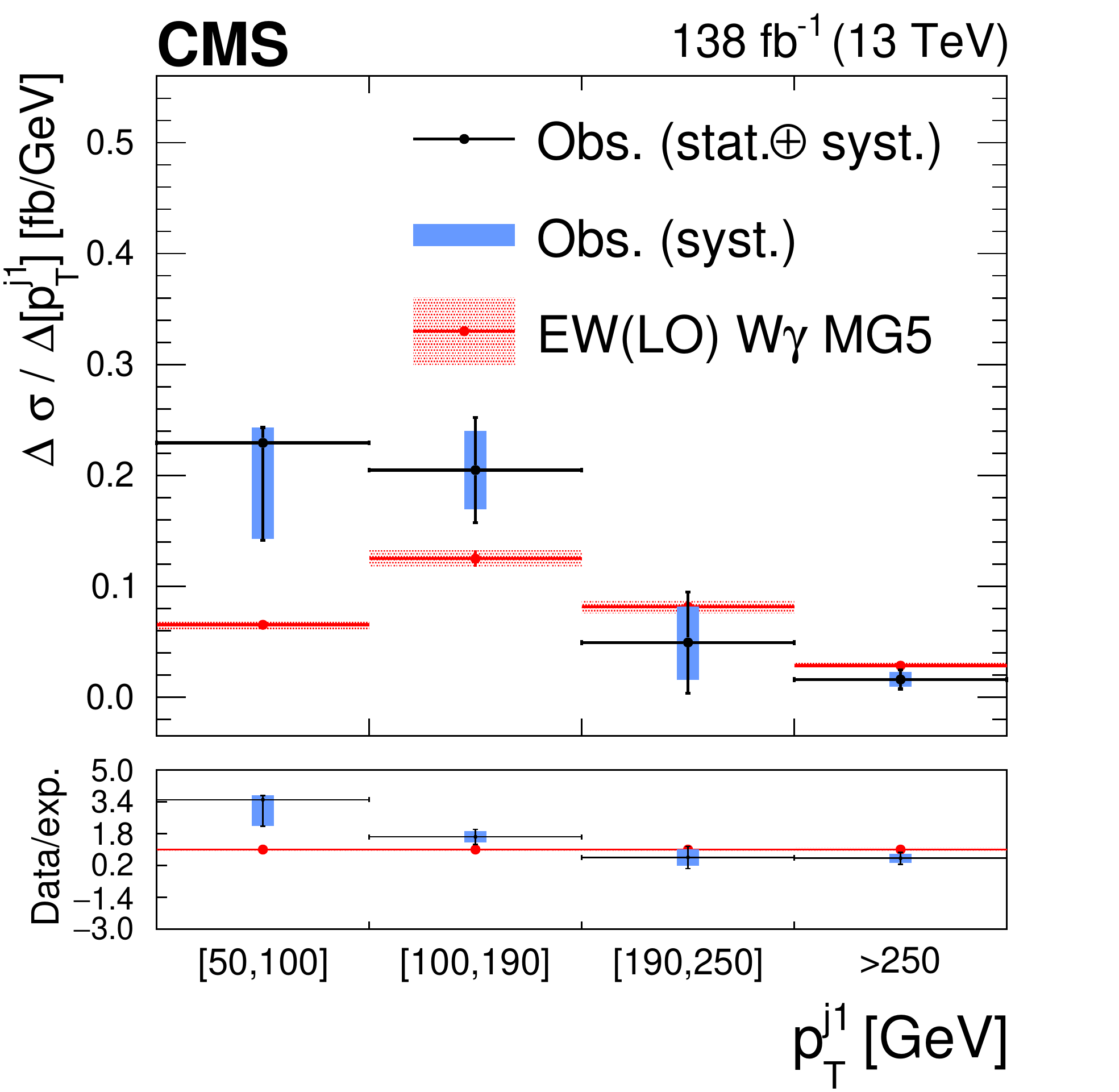}
      \\
      \includegraphics[width=0.40\textwidth]{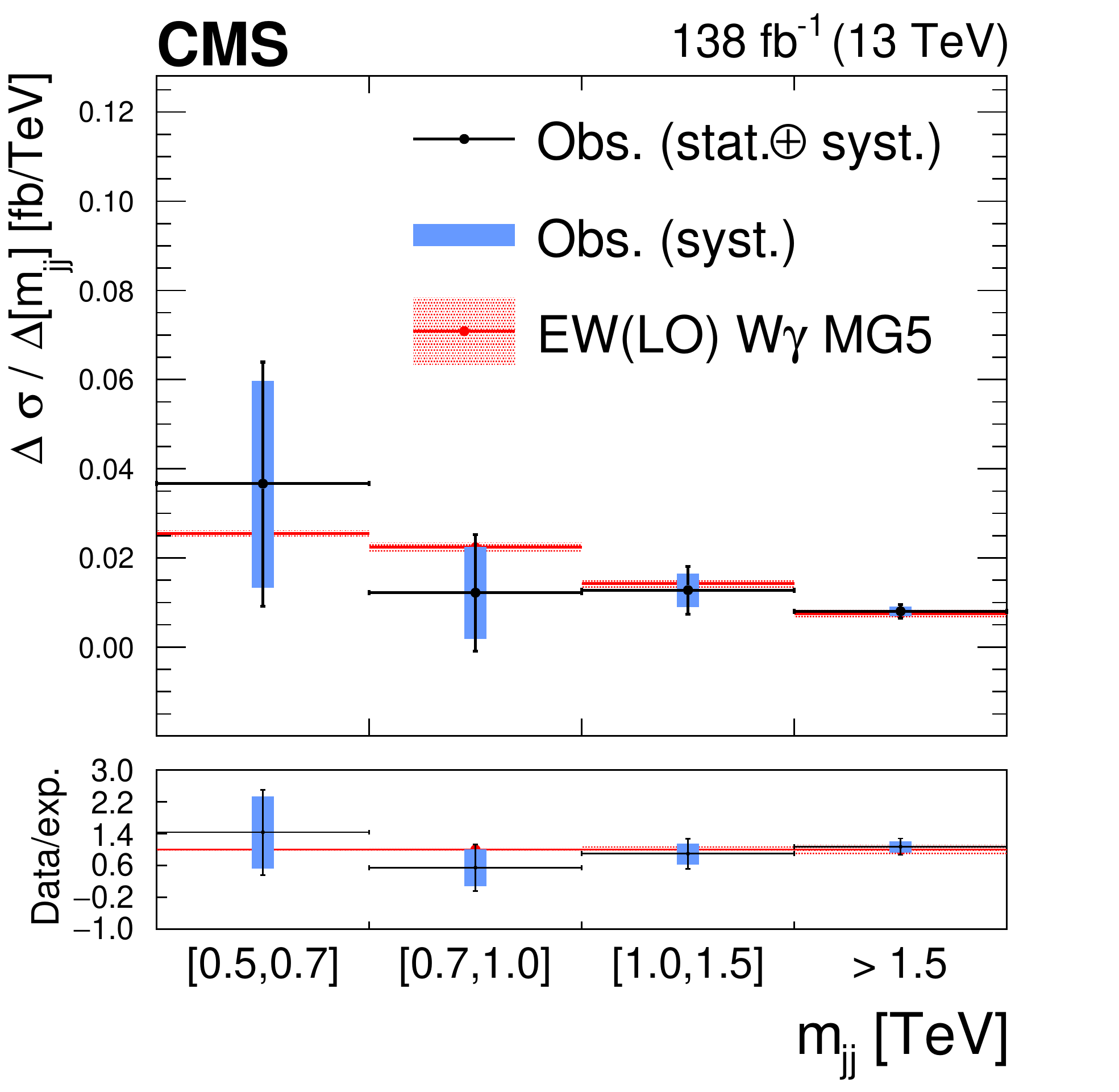}
      \includegraphics[width=0.40\textwidth]{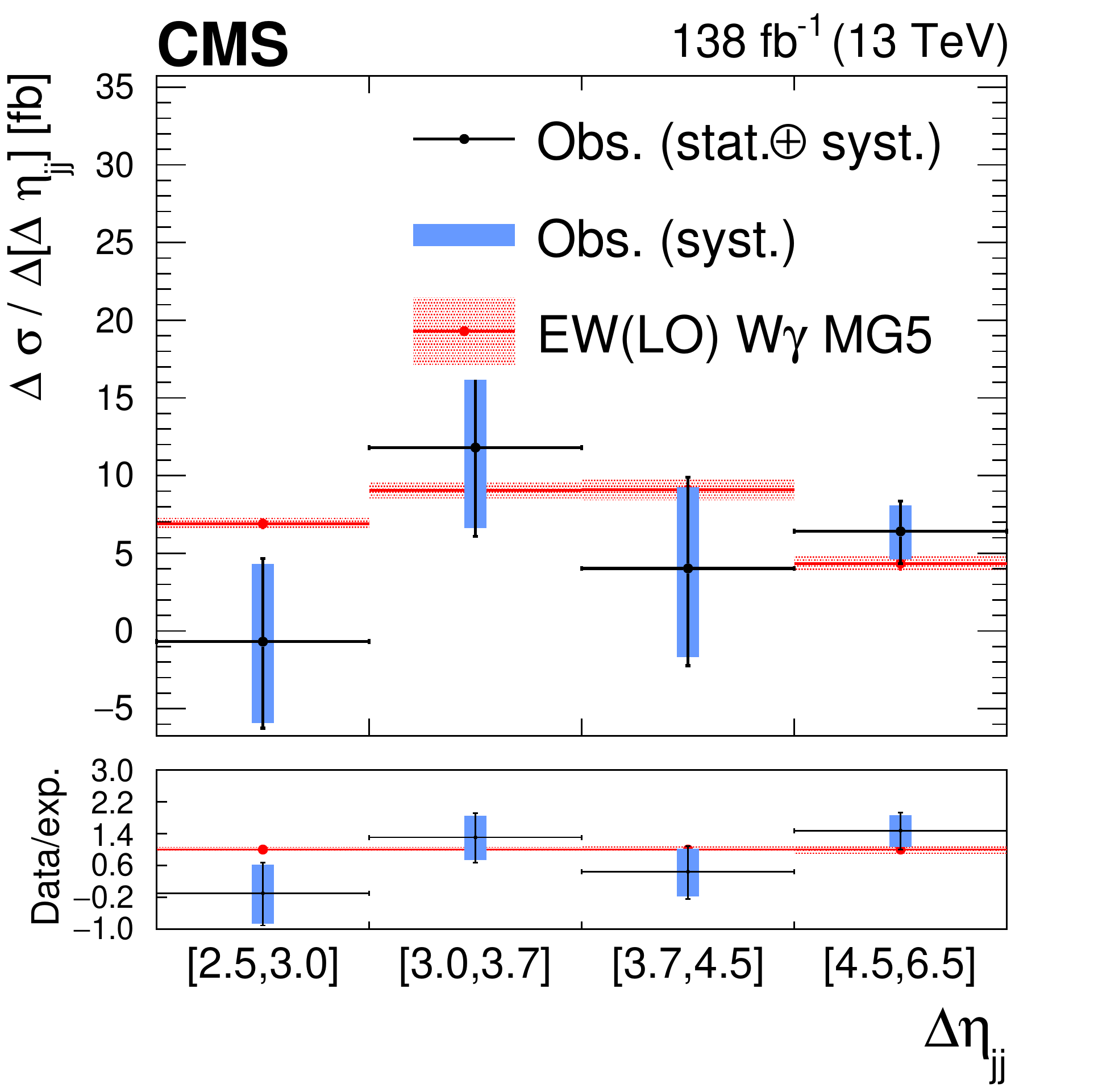}
      \caption{Differential cross sections for the EW {\Wgjj} production as functions of \ptX{\text{l}}, \ptX{\gamma}, \ptX{\text{j1}}, $m_{\ell\gamma}$, $m_{\text{jj}}$, and $\Delta\eta_{\text{jj}}$.
      The highest bins in each plot have no upper bound and are normalized by the bin boundaries of (110, 400), (170 200), (160, 1000), (250, 500), and (1500, 2000) \GeV for \ptX{\text{l}}, \ptX{\gamma}, $m_{\ell\gamma}$, \ptX{\text{j1}} and $m_{\text{jj}}$ respectively.
      The blue bands stand for the systematic uncertainties and the black bands represent the total uncertainties.}
      \label{fig:mig_mat_ew}
\end{figure*}

\begin{figure*}[htbp!]
   \centering
      \includegraphics[width=0.40\textwidth]{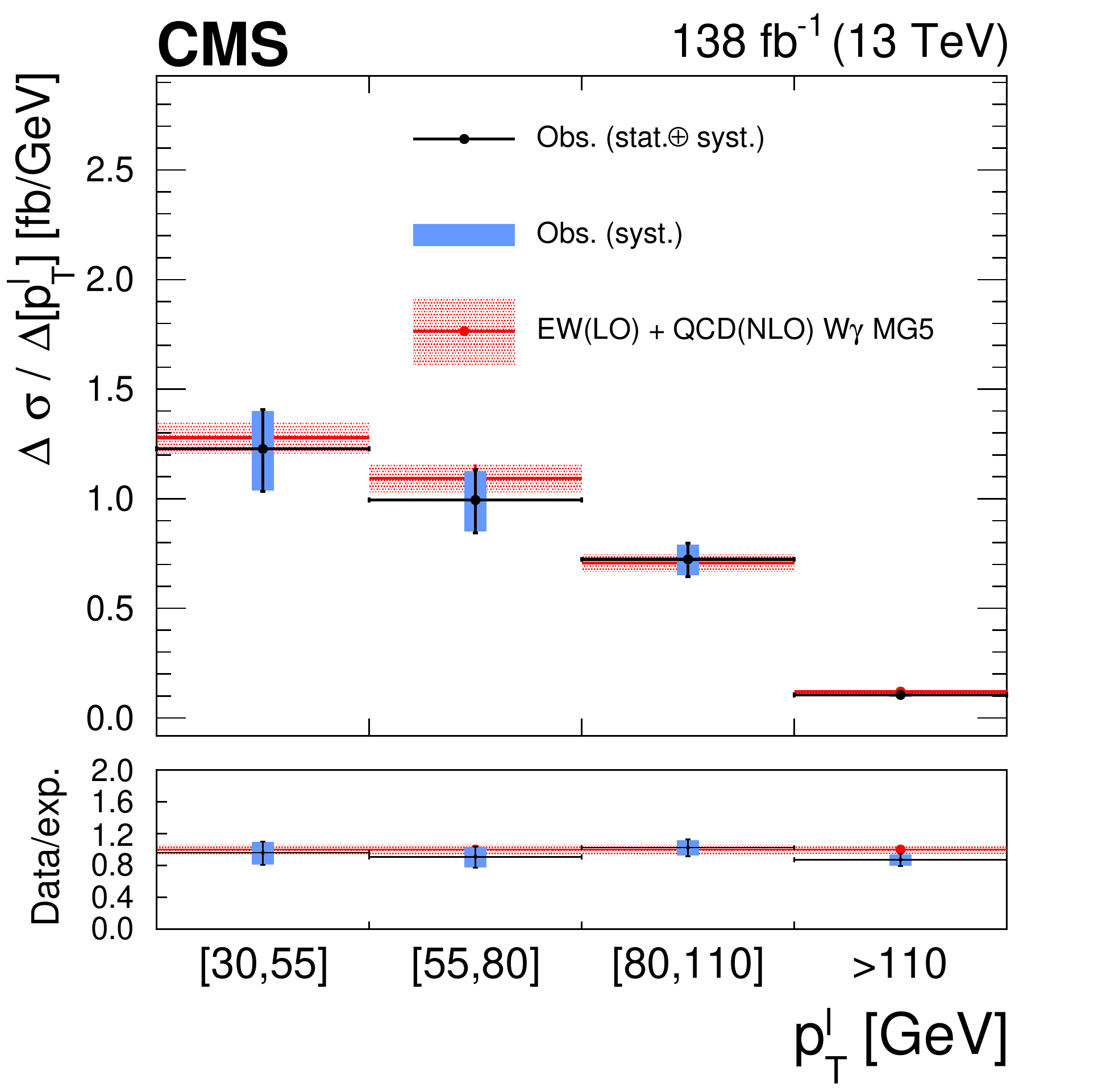}
      \includegraphics[width=0.40\textwidth]{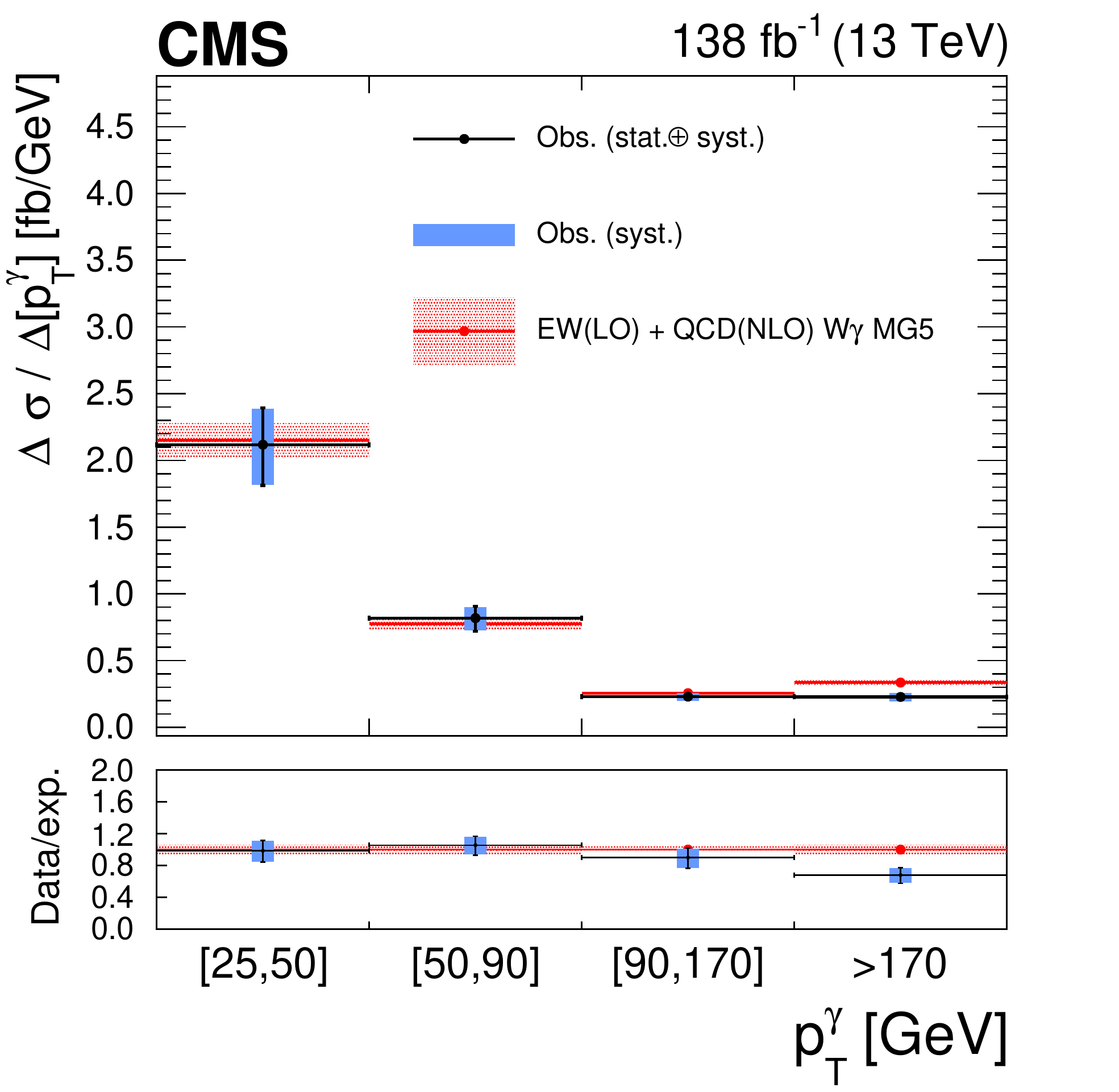}
      \\
      \includegraphics[width=0.40\textwidth]{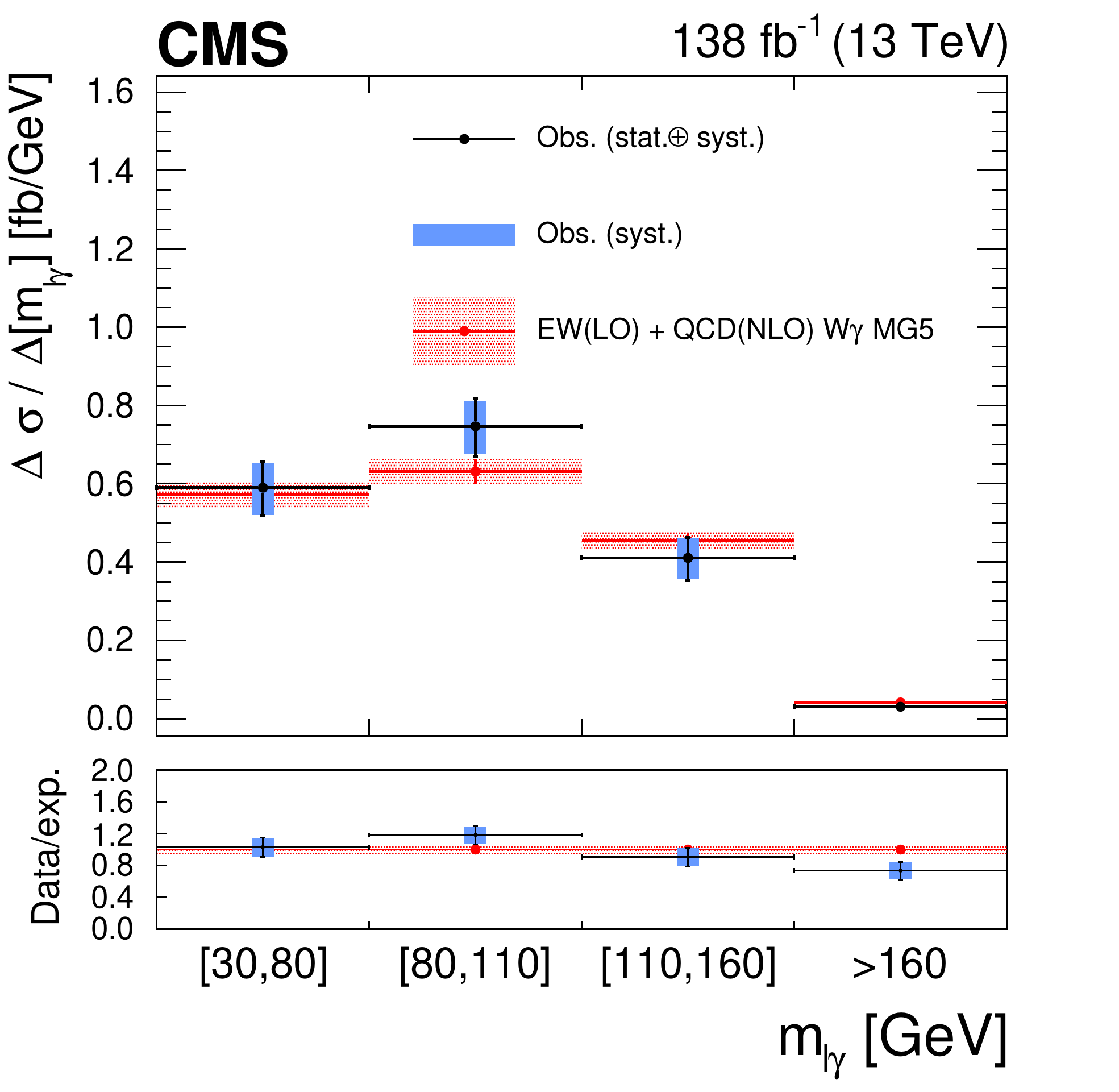}
      \includegraphics[width=0.40\textwidth]{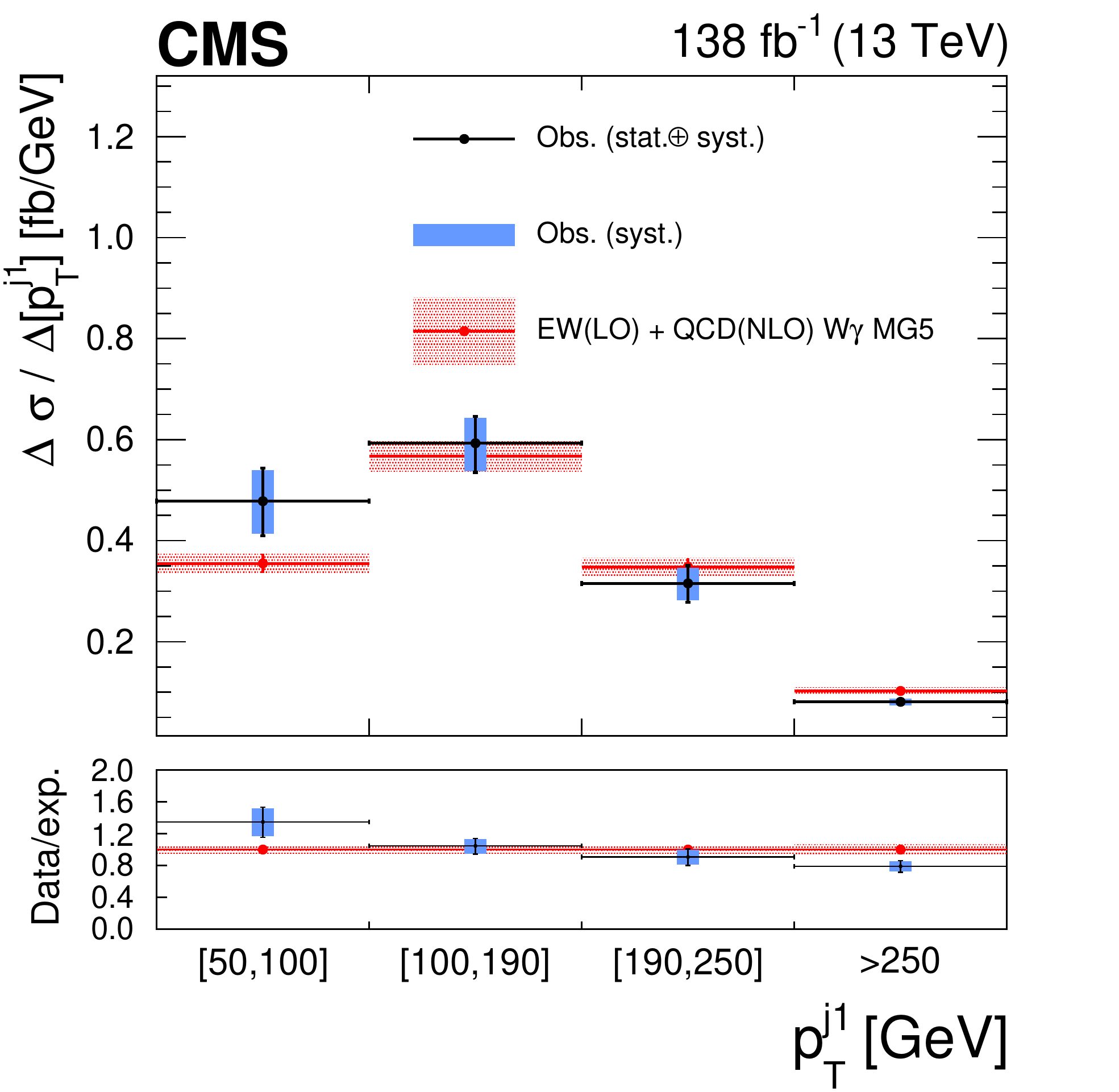}
      \\
      \includegraphics[width=0.40\textwidth]{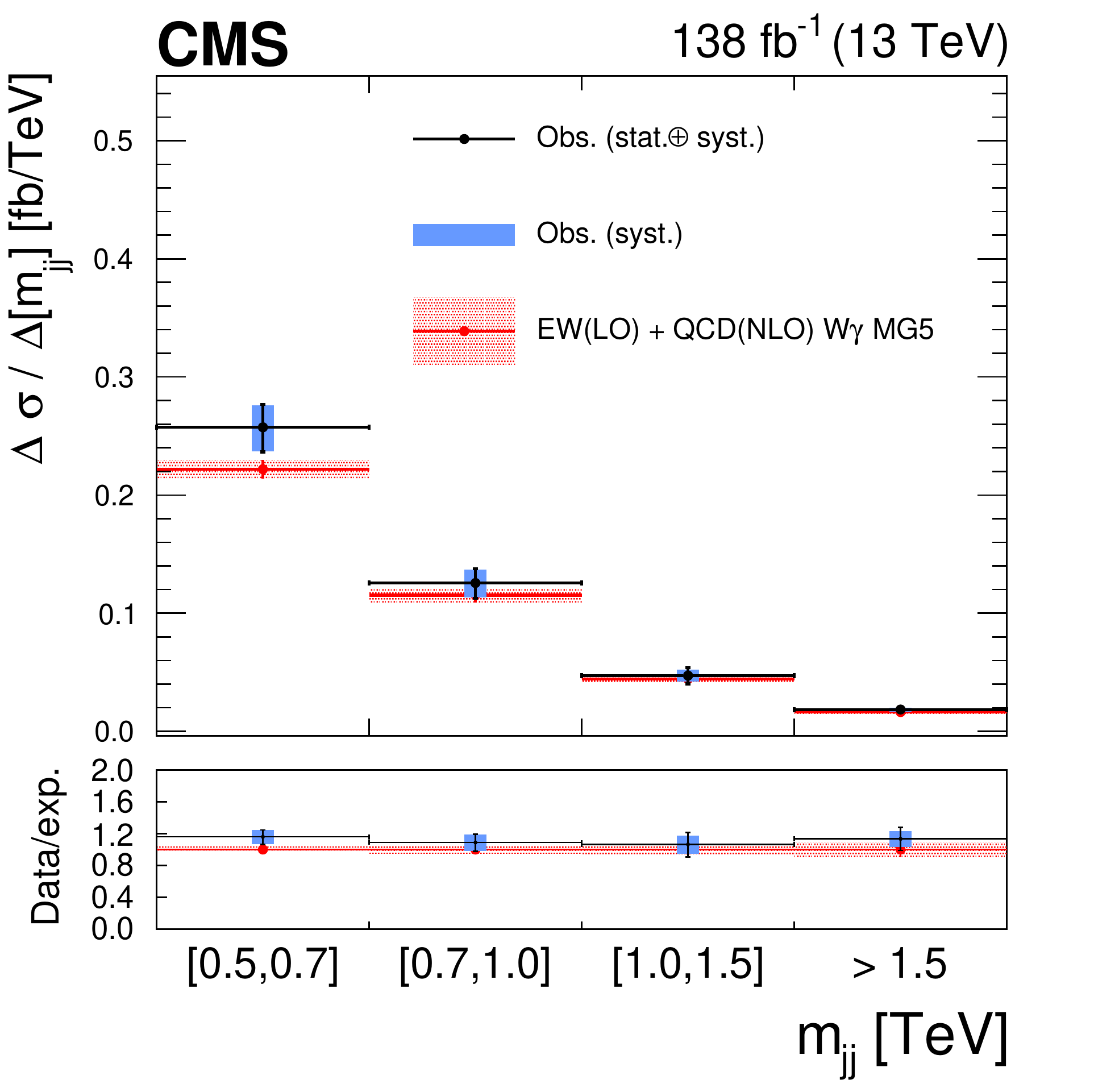}
      \includegraphics[width=0.40\textwidth]{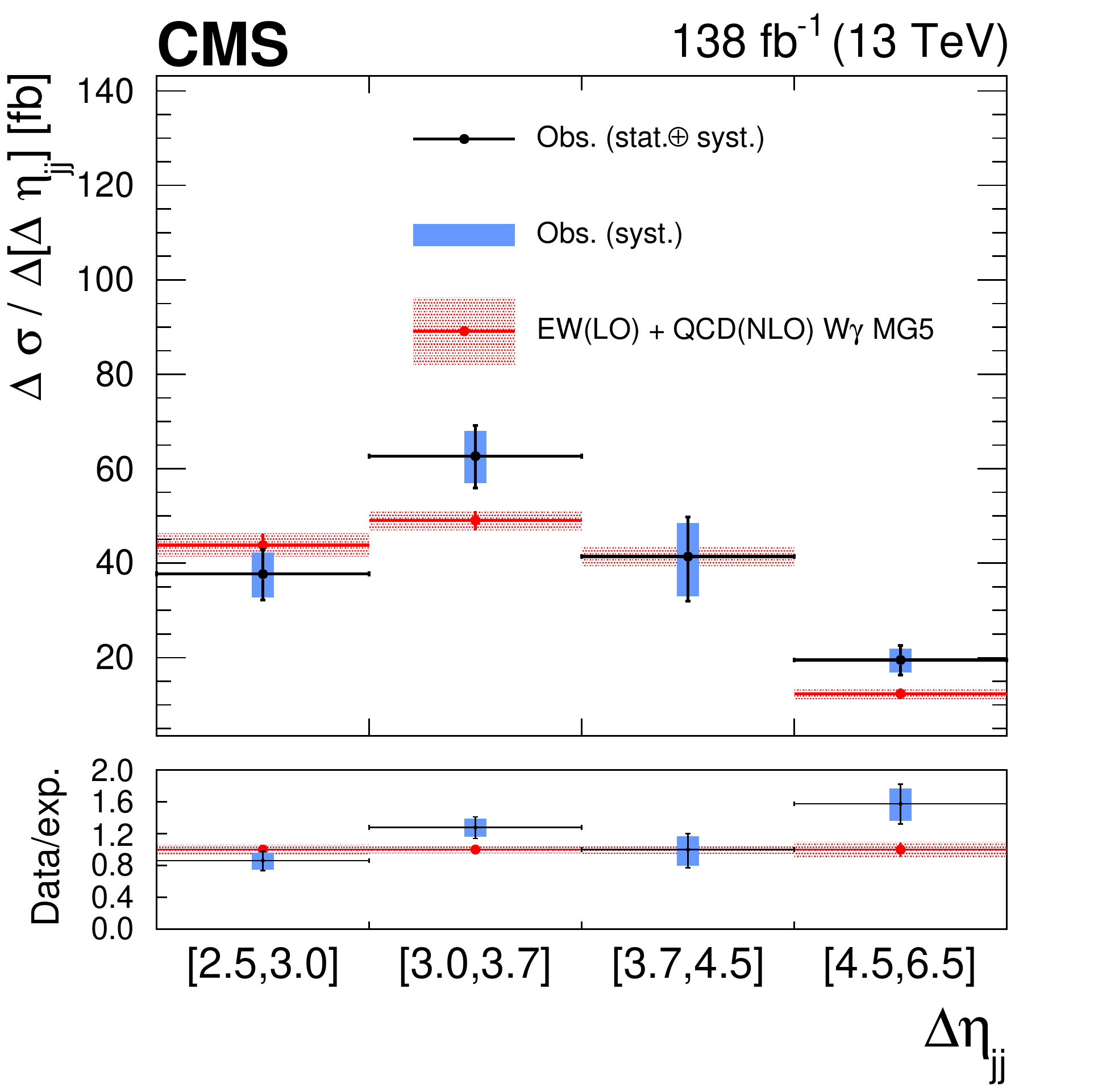}
     \caption{
      Differential cross sections for the EW+QCD {\Wgjj} production as functions of \ptX{\text{l}}, \ptX{\gamma}, \ptX{\text{j1}}, $m_{\ell\gamma}$, $m_{\text{jj}}$, and $\Delta\eta_{\text{jj}}$.
      The highest bins in each plot have no upper bound and are normalized by the bin boundaries of (110, 400), (170 200), (160, 1000), (250, 500), and (1500, 2000) \GeV for \ptX{\text{l}}, \ptX{\gamma}, $m_{\ell\gamma}$, \ptX{\text{j1}} and $m_{\text{jj}}$ respectively.
      The blue bands stand for the systematic uncertainties and the black bands represent the total uncertainties.
      }
      \label{fig:mig_mat_ewqcd}
\end{figure*}

The unfolded differential distributions are shown in Fig.~\ref{fig:mig_mat_ew} for the EW production and in Fig.~\ref{fig:mig_mat_ewqcd} for EW+QCD production.
Comparisons are shown with the theoretical predictions from MG5. The predictions are in agreement with the unfolded data in general.

\section{Limits on anomalous quartic gauge couplings}
\label{sec:aqgclimits}

{\tolerance=500
The effects of BSM physics can be parameterized in a generic way through
a set of linearly independent higher-dimensional operators in an effective field theory~\cite{paper_aqgc}.
As mentioned above, VBS is particularly suitable to constrain aQGCs. 
The lowest-dimension operators that modify quartic gauge couplings
but do not exhibit two or three weak gauge boson vertices are dimension-eight. 
Reference.~\cite{Eboli:2006wa} proposes nine independent charge-conjugate and parity-conserving 
dimension-eight effective operators by assuming the SU(2)$\times$U(1) symmetry of the EW gauge field.
The model includes a Higgs-field doublet to incorporate the presence of the SM Higgs boson.
The operators affecting the {\Wgjj} channel can be divided into two categories.
The operators \loperator{M}{0}--\loperator{M}{7} contain an SU(2) field strength,
the U(1) field strength, and the covariant derivative of the Higgs doublet field.
The operators \loperator{T}{0}--\loperator{T}{2} and \loperator{T}{5}--\loperator{T}{7}, contain only the two field strengths.
The coefficient of the operator \loperator{X}{Y} is denoted by \fcoefflam{X}{Y}, where $\Lambda$ is the unknown scale of BSM physics.
\par}

\begin{figure}[htbp!]
  \centering
      \includegraphics[width=0.47\textwidth]{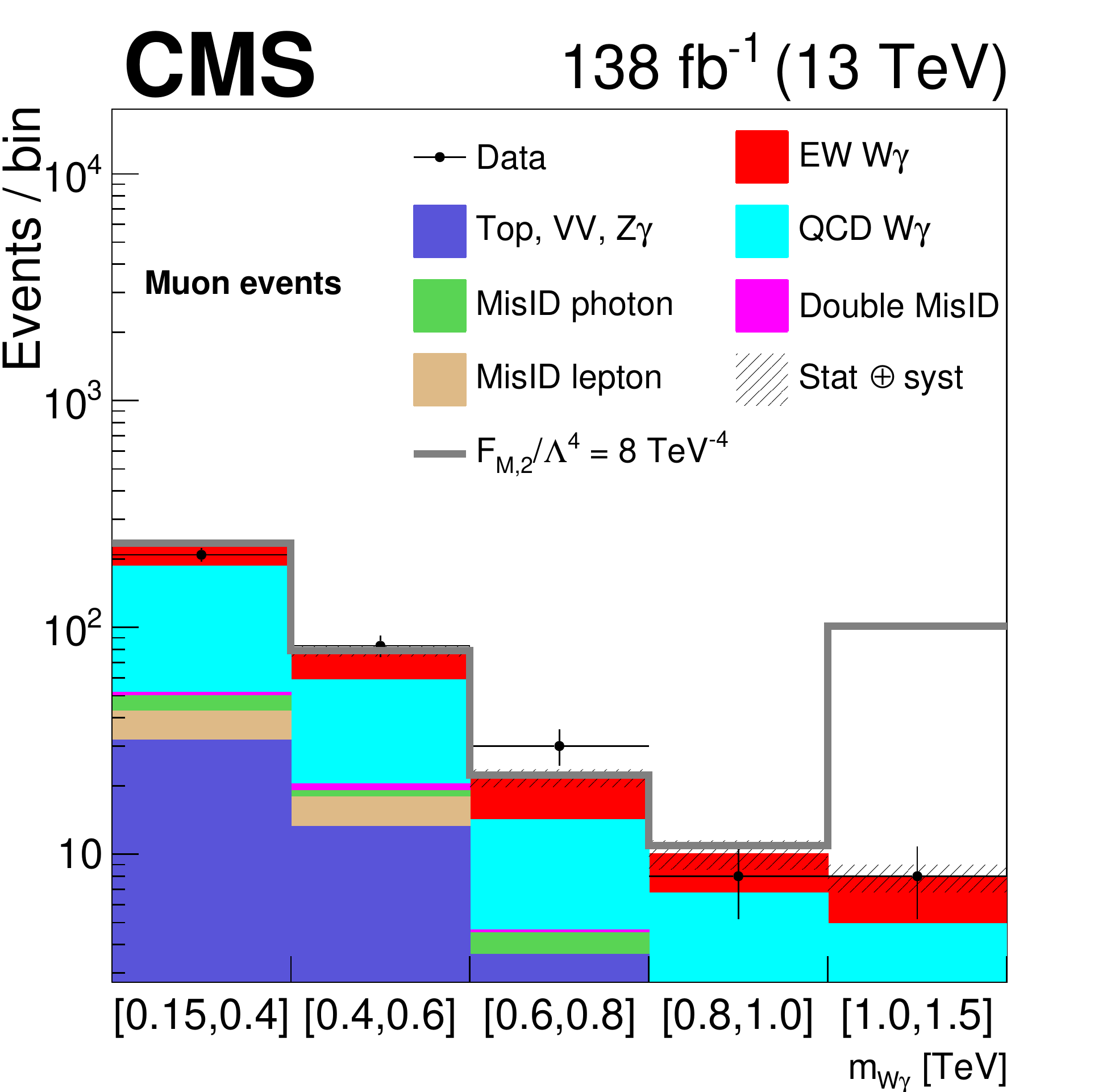}
      \includegraphics[width=0.47\textwidth]{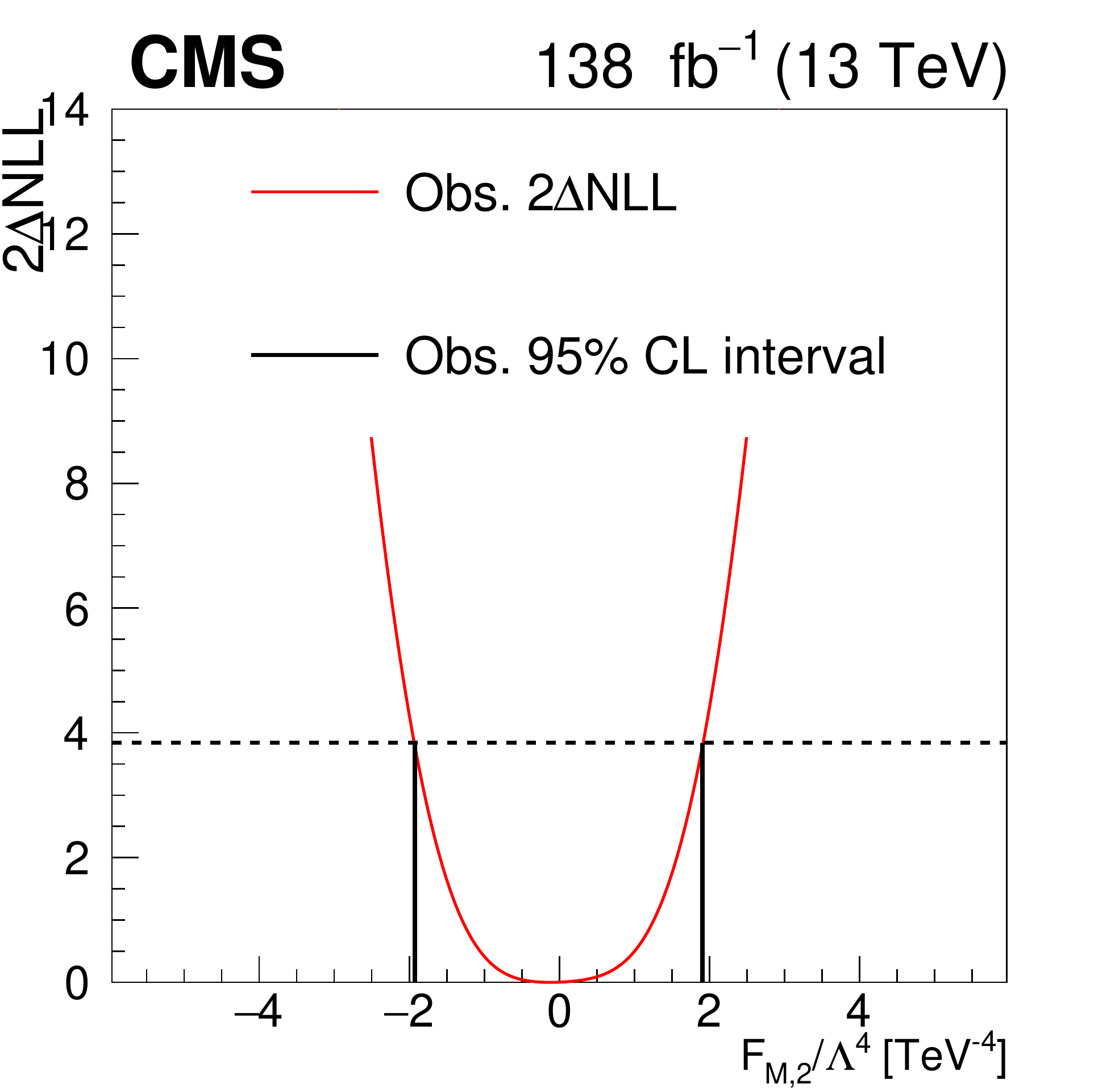}
      \caption{The \mWg distribution for muon events satisfying the aQGC region selection and used to set constraints on the anomalous gauge coupling parameters (\cmsLeft). Electron events, not shown here, are also used.
        The gray line represents a nonzero \fcoefflam{M}{2} setting.
        Events with $\mWg > 1500\GeV$ are included in the last bin.
        The hatched bands represent the combined statistical and systematic uncertainties on the predicted yields.
        The vertical bars on  the data points represent the statistical uncertainties of data.
        Likelihood scan and the observed 95\% \CL interval for the aQGC parameter \fcoefflam{M}{2} (\cmsRight).}
      \label{fig:aqgc_mu}
\end{figure}

A simulation is performed that includes the effects of aQGCs in addition to the SM EW {\Wgjj} production,
as well as the interference between the two contributions.
Since a contribution from aQGCs would enhance the production of events with large {\Wg} mass, we therefore use this observable to extract limits on the aQGC parameters.
To obtain a continuous prediction for the signal as a function of each anomalous coupling,
a quadratic fit is performed to the SM+aQGC yield as a function of the aQGC coefficient value,
separately in each \mWg bin.
In addition to the selection described in Section~\ref{sec:selection},
further requirements are applied to exploit the fact that the aQGC contributions
arise from pure VBS diagrams, and are thus enhanced in the VBS phase space region,
and the anomalous operators lead to more energetic final-state particles.
These requirements are optimized to enhance the aQGC sensitivity, based on simulation studies,
and are: $\mjj > 800\GeV$, $\abs{\dejj} > 2.5$, $\mWg > 150\GeV$, and $\ptg > 100\GeV$.
As an example, Fig.~\ref{fig:aqgc_mu} (\cmsLeft) shows the resulting \mWg distribution in muon channel.

We set two-sided limits on the operator coefficients through a limit-setting procedure
that involves first obtaining the global maximum of the profile likelihood function,
and then the maximum of the profile likelihood function at fixed coefficient values,
which are compared with the global maximum and converted to confidence level (\CL) intervals.
Figure~\ref{fig:aqgc_mu} (\cmsRight) shows the likelihood scan for the \fcoefflam{M}{2} parameter in the calculation of the observed limits.

The observed and expected 95\% \CL limits on the aQGC coefficients are summarized in Tab.~\ref{tab:VBS_aQGC_3years}. These are the most stringent limits to date on the aQGC parameters \fcoefflam{M}{2--5} and \fcoefflam{T}{6--7}. 

\begin{table}[ht!]
\centering
\topcaption{Exclusion limits at the 95\% \CL for each aQGC coefficient, derived from the \mWg distribution, assuming all other coefficients are set to zero. Unitarity bounds corresponding to each operator are also listed.
    All coupling parameter limits are in ${\TeVns}^{-4}$, while \ubound values are in \TeV.}
\begin{scotch*}{cccc}
Expected limit & Observed limit & \ubound \\
\hline
         -5.1  $< f_{{M,0}}/\Lambda^{4} <$ 5.1 & $ -5.6  < f_{{M,0}}/\Lambda^{4} < 5.5  $ & 1.7  \\
         -7.1  $< f_{{M,1}}/\Lambda^{4} <$ 7.4 & $ -7.8 < f_{{M,1}}/\Lambda^{4} < 8.1 $ & 2.1 \\
         -1.8  $< f_{{M,2}}/\Lambda^{4} <$ 1.8 & $ -1.9  < f_{{M,2}}/\Lambda^{4} < 1.9  $ & 2.0  \\
         -2.5  $< f_{{M,3}}/\Lambda^{4} <$ 2.5 & $ -2.7  < f_{{M,3}}/\Lambda^{4} < 2.7  $ & 2.7  \\
         -3.3  $< f_{{M,4}}/\Lambda^{4} <$ 3.3 & $ -3.7  < f_{{M,4}}/\Lambda^{4} < 3.6  $ & 2.3  \\
         -3.4  $< f_{{M,5}}/\Lambda^{4} <$ 3.6 & $ -3.9  < f_{{M,5}}/\Lambda^{4} < 3.9  $ & 2.7  \\
         -13  $< f_{{M,7}}/\Lambda^{4} <$ 13 & $ -14 < f_{{M7}}/\Lambda^{4} < 14 $ & 2.2  \\
         -0.43  $< f_{{T,0}}/\Lambda^{4} <$ 0.51 & $ -0.47  < f_{{T,0}}/\Lambda^{4} < 0.51  $ & 1.9  \\
         -0.27  $< f_{{T,1}}/\Lambda^{4} <$ 0.31 & $ -0.31  < f_{{T,1}}/\Lambda^{4} < 0.34  $ & 2.5  \\
         -0.72  $< f_{{T,2}}/\Lambda^{4} <$ 0.92 & $ -0.85  < f_{{T,2}}/\Lambda^{4} < 1.0  $ & 2.3  \\
         -0.29  $< f_{{T,5}}/\Lambda^{4} <$ 0.31 & $ -0.31  < f_{{T,5}}/\Lambda^{4} < 0.33  $ & 2.6  \\
         -0.23  $< f_{{T,6}}/\Lambda^{4} <$ 0.25 & $ -0.25  < f_{{T,6}}/\Lambda^{4} < 0.27  $ & 2.9  \\
         -0.60  $< f_{{T,7}}/\Lambda^{4} <$ 0.68 & $ -0.67  < f_{{T,7}}/\Lambda^{4} < 0.73  $ & 3.1  \\

\end{scotch*}
\label{tab:VBS_aQGC_3years}
\end{table}

They are obtained by varying the coefficient of one operator at a time, with all others set to zero, \ie, the SM value.
The yield of the EW signal in any bin is a quadratic function of the coefficient,
whose minimum in general does not occur at a coefficient value of zero because of the interference with the SM operators. The constraints set on the aQGCs are compatible with the SM predictions of zero.
The NLO EW corrections to VBS \Wg can be sizable and increase as a function of \mjj,
which may bias the aQGC measurement. Although there is no NLO EW calculation available yet for VBS \Wg production, we have checked, using the numbers from same-sign WW scattering~\cite{Biedermann:2016yds,Biedermann:2017bss},
that the effect on the aQGC limits is negligible.
The unitarity bound (\ubound) is defined as the scattering energy at which the aQGC coupling strength,
when set equal to the observed limit, would result in a scattering amplitude that violates unitarity.
The value of \ubound is determined using the analytical formulas from Ref.~\cite{Almeida:2020ylr}.

\section{Summary}
\label{sec:summary}
Measurements of the electroweak (EW) production of a \PW boson, a photon,
and two jets in proton-proton collisions at a center-of-mass energy of 13\TeV have been presented. 
The data correspond to an integrated luminosity of 138\fbinv in Run 2 collected with the CMS detector. 
Events are selected by requiring one isolated lepton (electron or muon) with high transverse momentum (\pt),
a moderate missing transverse momentum, one high-\pt isolated photon,
and two jets with a large rapidity separation and a large dijet mass.
The signal is observed for the first time at 13\TeV, with an observed (expected) significance of 6.0 (6.8) standard deviations, where the expectation is based on the standard model predictions.
In a restricted fiducial region, the cross section for the EW \Wgjj production is $23.5^{+4.9}_{-4.7}\unit{fb}$ and the cross section for the total EW+QCD \Wgjj production is $113\pm 13\unit{fb}$.
Both measurements are consistent with standard model predictions. 
For the first time, differential cross sections for EW \Wgjj and for EW+QCD \Wgjj production are measured.
Constraints placed on anomalous quartic gauge couplings in terms of
dimension-8 effective field theory operators are extracted and are the most stringent limits to date on the aQGC parameters \fcoefflam{M}{2--5} and \fcoefflam{T}{6--7}. 

\begin{acknowledgments}
We congratulate our colleagues in the CERN accelerator departments for the excellent performance of the LHC and thank the technical and administrative staffs at CERN and at other CMS institutes for their contributions to the success of the CMS effort. In addition, we gratefully acknowledge the computing centers and personnel of the Worldwide LHC Computing Grid and other centers for delivering so effectively the computing infrastructure essential to our analyses. Finally, we acknowledge the enduring support for the construction and operation of the LHC, the CMS detector, and the supporting computing infrastructure provided by the following funding agencies: BMBWF and FWF (Austria); FNRS and FWO (Belgium); CNPq, CAPES, FAPERJ, FAPERGS, and FAPESP (Brazil); MES and BNSF (Bulgaria); CERN; CAS, MoST, and NSFC (China); MINCIENCIAS (Colombia); MSES and CSF (Croatia); RIF (Cyprus); SENESCYT (Ecuador); MoER, ERC PUT and ERDF (Estonia); Academy of Finland, MEC, and HIP (Finland); CEA and CNRS/IN2P3 (France); BMBF, DFG, and HGF (Germany); GSRI (Greece); NKFIH (Hungary); DAE and DST (India); IPM (Iran); SFI (Ireland); INFN (Italy); MSIP and NRF (Republic of Korea); MES (Latvia); LAS (Lithuania); MOE and UM (Malaysia); BUAP, CINVESTAV, CONACYT, LNS, SEP, and UASLP-FAI (Mexico); MOS (Montenegro); MBIE (New Zealand); PAEC (Pakistan); MES and NSC (Poland); FCT (Portugal); MESTD (Serbia); MCIN/AEI and PCTI (Spain); MOSTR (Sri Lanka); Swiss Funding Agencies (Switzerland); MST (Taipei); MHESI and NSTDA (Thailand); TUBITAK and TENMAK (Turkey); NASU (Ukraine); STFC (United Kingdom); DOE and NSF (USA).
    
\hyphenation{Rachada-pisek} Individuals have received support from the Marie-Curie program and the European Research Council and Horizon 2020 Grant, contract Nos.\ 675440, 724704, 752730, 758316, 765710, 824093, 884104, and COST Action CA16108 (European Union); the Leventis Foundation; the Alfred P.\ Sloan Foundation; the Alexander von Humboldt Foundation; the Belgian Federal Science Policy Office; the Fonds pour la Formation \`a la Recherche dans l'Industrie et dans l'Agriculture (FRIA-Belgium); the Agentschap voor Innovatie door Wetenschap en Technologie (IWT-Belgium); the F.R.S.-FNRS and FWO (Belgium) under the ``Excellence of Science -- EOS" -- be.h project n.\ 30820817; the Beijing Municipal Science \& Technology Commission, No. Z191100007219010; the Ministry of Education, Youth and Sports (MEYS) of the Czech Republic; the Hellenic Foundation for Research and Innovation (HFRI), Project Number 2288 (Greece); the Deutsche Forschungsgemeinschaft (DFG), under Germany's Excellence Strategy -- EXC 2121 ``Quantum Universe" -- 390833306, and under project number 400140256 - GRK2497; the Hungarian Academy of Sciences, the New National Excellence Program - \'UNKP, the NKFIH research grants K 124845, K 124850, K 128713, K 128786, K 129058, K 131991, K 133046, K 138136, K 143460, K 143477, 2020-2.2.1-ED-2021-00181, and TKP2021-NKTA-64 (Hungary); the Council of Science and Industrial Research, India; the Latvian Council of Science; the Ministry of Education and Science, project no. 2022/WK/14, and the National Science Center, contracts Opus 2021/41/B/ST2/01369 and 2021/43/B/ST2/01552 (Poland); the Funda\c{c}\~ao para a Ci\^encia e a Tecnologia, grant CEECIND/01334/2018 (Portugal); the National Priorities Research Program by Qatar National Research Fund; MCIN/AEI/10.13039/501100011033, ERDF ``a way of making Europe", and the Programa Estatal de Fomento de la Investigaci{\'o}n Cient{\'i}fica y T{\'e}cnica de Excelencia Mar\'{\i}a de Maeztu, grant MDM-2017-0765 and Programa Severo Ochoa del Principado de Asturias (Spain); the Chulalongkorn Academic into Its 2nd Century Project Advancement Project, and the National Science, Research and Innovation Fund via the Program Management Unit for Human Resources \& Institutional Development, Research and Innovation, grant B05F650021 (Thailand); the Kavli Foundation; the Nvidia Corporation; the SuperMicro Corporation; the Welch Foundation, contract C-1845; and the Weston Havens Foundation (USA).
\end{acknowledgments}

\bibliography{auto_generated}
\cleardoublepage \appendix\section{The CMS Collaboration \label{app:collab}}\begin{sloppypar}\hyphenpenalty=5000\widowpenalty=500\clubpenalty=5000
\cmsinstitute{Yerevan Physics Institute, Yerevan, Armenia}
{\tolerance=6000
A.~Tumasyan\cmsAuthorMark{1}\cmsorcid{0009-0000-0684-6742}
\par}
\cmsinstitute{Institut f\"{u}r Hochenergiephysik, Vienna, Austria}
{\tolerance=6000
W.~Adam\cmsorcid{0000-0001-9099-4341}, J.W.~Andrejkovic, T.~Bergauer\cmsorcid{0000-0002-5786-0293}, S.~Chatterjee\cmsorcid{0000-0003-2660-0349}, K.~Damanakis\cmsorcid{0000-0001-5389-2872}, M.~Dragicevic\cmsorcid{0000-0003-1967-6783}, A.~Escalante~Del~Valle\cmsorcid{0000-0002-9702-6359}, P.S.~Hussain\cmsorcid{0000-0002-4825-5278}, M.~Jeitler\cmsAuthorMark{2}\cmsorcid{0000-0002-5141-9560}, N.~Krammer\cmsorcid{0000-0002-0548-0985}, L.~Lechner\cmsorcid{0000-0002-3065-1141}, D.~Liko\cmsorcid{0000-0002-3380-473X}, I.~Mikulec\cmsorcid{0000-0003-0385-2746}, P.~Paulitsch, F.M.~Pitters, J.~Schieck\cmsAuthorMark{2}\cmsorcid{0000-0002-1058-8093}, R.~Sch\"{o}fbeck\cmsorcid{0000-0002-2332-8784}, D.~Schwarz\cmsorcid{0000-0002-3821-7331}, M.~Sonawane\cmsorcid{0000-0003-0510-7010}, S.~Templ\cmsorcid{0000-0003-3137-5692}, W.~Waltenberger\cmsorcid{0000-0002-6215-7228}, C.-E.~Wulz\cmsAuthorMark{2}\cmsorcid{0000-0001-9226-5812}
% ...
\par}
\vskip\cmsinstskip
\dag:~Deceased\\
$^{1}$Also at Yerevan State University, Yerevan, Armenia\\
$^{2}$Also at TU Wien, Vienna, Austria\\
% ...
\end{sloppypar}
%%% END EDITABLE REGION %%%
% skeleton_end
\end{document}